%% file: SU12Model10C.tex
\documentclass[preprint,showpacs,amsfonts,amssymb,amsmath,aps,prd]{revtex4}
\usepackage[latin1]{inputenc}
\usepackage[T1]{fontenc}
\usepackage[english]{babel}
\usepackage[pdftex]{xcolor}
\usepackage{graphicx}
\usepackage{xspace,tabularx}

\definecolor{darkred}{rgb}{0.5,0,0}
\definecolor{darkgreen}{rgb}{0,.5,0}
\definecolor{darkblue}{rgb}{0,0,.5}
\usepackage[pdftex,bookmarksopen]{hyperref}
\hypersetup{
  pdftitle={Unification of gauge, family, and flavor symmetries illustrated in 
gauged SU(12) models}, %%
  pdfauthor={Robert Feger}, %%
  pdfsubject={}, %%
  pdfcreator={pdfTeX},
  pdfproducer={Robert Feger}, %%
  pdfkeywords={},%%
  colorlinks=true,        % false: boxed links; true: colored links
  linkcolor=darkblue,      % color of internal links
  citecolor=darkblue,    % color of links to bibliography
  urlcolor=darkblue,      % color of external links
  pdfborder={0 0 0}
}

\renewcommand{\epsilon}{\varepsilon}
\DeclareMathOperator{\diag}{diag}

\newcommand\abs[1]{\lvert#1\rvert}
\newcommand{\SU}[1]{\ensuremath{{\text{SU}(#1)}}}
\newcommand{\U}[1]{\ensuremath{{\text{U}(#1)}}}

\newcommand{\irrep}[1]{\ensuremath{\boldirrep{#1}}}

\newcommand{\boldirrep}{\mathbf}
\newlength{\irrepwidth}
\newlength{\irrepbarthickness}
\newlength{\irrepbarheight}
\newcommand{\irrepbar}[1]{%
	\settoheight{\irrepbarheight}{\ensuremath{\boldirrep{#1}}}%
	\settowidth{\irrepwidth}{\ensuremath{\boldirrep{#1}}}%
	\makebox[0pt][l]{\ensuremath{\boldirrep{#1}}}%
	\rule[1.2\irrepbarheight]{\irrepwidth}{\irrepbarthickness}%
}

\newcommand{\irrepsub}[2]{\ensuremath{\irrep{#1}_\text{#2}}}
\newcommand{\irrepbarsub}[2]{\ensuremath{\irrepbar{#1}_\text{#2}}}

\newcommand\fermion\irrepsub
\newcommand\fermionbar\irrepbarsub

\newcommand{\higgs}[1]{\irrepsub{#1}{H}}
\newcommand{\higgsbar}[1]{\irrepbarsub{#1}{H}}

\newcommand{\massivefermionpair}[2]{%
\ensuremath{#1{\times}#2}%
}
\newcommand{\UpType}[2]{\text{\textbf{U#1#2}:}}
\newcommand{\DownType}[2]{\text{\textbf{D#1#2}:}}
\newcommand{\Majorana}[2]{\text{\textbf{MN#1#2}:}}
\newcommand{\Dirac}[2]{\text{\textbf{DN#1#2}:}}

\newcommand\hu[1]{h^\text{u}_{#1}}
\newcommand\hd[1]{h^\text{d}_{#1}}
\newcommand\hl[1]{h^\ell_{#1}}
\newcommand\hmn[1]{{h^\text{mn}_{#1}}}
\newcommand\hdn[1]{{h^\text{dn}_{#1}}}

\begin{document}
\begin{flushright}
    FERMILAB-PUB-16-014-T
\end{flushright}
\vspace{0.2in}
\title{\boldmath%
    Unification of gauge, family, and flavor symmetries\\ 
    illustrated in gauged \SU{12} models}

\date{February 25, 2016\\[0.2in]}

\author{\bf Carl H. Albright$^{1,2,*}$, Robert P. Feger$^{3,\dagger}$, and Thomas
  W.  Kephart$^{4,\ddagger}$}
\affiliation{$^1$Department of Physics, Northern Illinois University,
  DeKalb, IL 60115\\
  $^2$Theoretical Physics, Fermilab, Batavia, IL 60510\\
  $^3$Institut f\"{u}r Theoretische Physik und Astrophysik,\\ 
    Universit\"{a}t W\"{u}rzburg, Am Hubland, D-97074 W\"{u}rzburg, Germany\\
  $^4$Department of Physics and Astronomy, Vanderbilt University,
  Nashville, TN 37235}
%\vspace{0.3in}
\begin{abstract}
To explain quark and lepton masses and mixing angles, one has to extend the 
standard model, and the usual practice is to put the quarks and leptons into 
irreducible representations of discrete groups. We argue that discrete flavor 
symmetries (and their concomitant problems) can be avoided if we extend the 
gauge group.  In the framework of \SU{12} we give explicit examples of 
models having varying degrees of predictability obtained by scanning 
over groups and representations and identifying cases with operators 
contributing to mass and mixing matrices that need little fine-tuning of 
prefactors. Fitting with quark and lepton masses run to the GUT scale and 
known mixing angles allows us to make predictions for the neutrino masses and 
hierarchy, the octant of the atmospheric mixing angle, leptonic ${\rm CP}$ 
violation, Majorana phases, and the effective mass observed in neutrinoless 
double beta decay.
\end{abstract}
% \keywords{family unification; fermion mass hierarchy}
\pacs{%
12.10.Dm, % Unified theories and models of strong and electroweak interactions
%,12.10.Kt % Unification of couplings; mass relations
12.15.Ff,
% Quark and lepton masses and mix
%  (see also 14.60.Pq Neutrino mass and mixing)
%    ,12.60.Fr % Extensions of electroweak Higgs sector
14.60.Pq \\% Neutrino mass and mixing
%(see also 12.15.Ff Quark and lepton masses and mixing)
%    ,14.60.St % Non-standard-model neutrinos, right-handed neutrinos, etc.
\\[-0.9in]}

\email[]{{albright@fnal.gov}\\[-0.1in]}
\email[$^\dagger$]{{robert.feger@gmail.com}\\[-0.1in]}
\email[$^\ddagger$]{{thomas.w.kephart@vanderbilt.edu}}

\maketitle
\thispagestyle{empty}

% \listoftodos

\section{INTRODUCTION}
\label{sec:Introduction}
Family and flavor symmetries of the observed quarks and leptons appear to be 
intimately related and remain much of a mystery today as to their precise 
structures. 
Although there is some ambiguity in the literature, here we make use of family 
symmetry to relate particles within a family of quarks and leptons as in 
the standard model (SM) or within some grand unified symmetry (GUTs) such as
\SU{5}, SO(10) or ${\rm E}_6$.  Flavor symmetry, on the other
hand, relates families which appear to be replicas of each other.  The flavor
symmetry may be continuous as in the case of \SU{3}, \SU{2}, \U{1} or 
discrete as in the case of ${\rm Z_2,\ Z_2 \times Z_2,\ S_3,\ A_4,\ S_4}$, 
etc. (For reviews see \cite{Altarelli:2010gt,Ishimori:2010au,King:2013}.) The 
conventional picture is to assume a direct product symmetry 
group, $\mathbf{G_{\rm family} \times G_{\rm flavor}}$, where ${\bf G}_{\rm family}$
is gauged but ${\bf G}_{\rm flavor}$ is discrete. The necessity of 
${\bf G}_{\rm flavor}$ reflects the replication of families due to 
the fact that there are too few chiral, exotic free, irreducible 
representations (irreps) 
in the family groups for the observed chiral fermion families: just 
$\mathbf{\bar{5}}$ and $\mathbf{10}$ for \SU{5}, $\mathbf{16}$ for SO(10), 
and $\mathbf{27}$~for~${\rm E}_6$.  

Family and flavor unification requires a higher rank simple group. Some early
attempts were based on \SU{11}, \SU{8}, \SU{9} and SO(18),~\cite{Georgi:1979md,
PHF,Frampton:1979fd,Frampton:2009ce,Fujimoto:1981bv} but none were
completely satisfactory \cite{Terazawa:1976xx}.  More recently such 
unification has been proposed in the framework of string compactification, 
see \cite{Kim:2015}.  Here we describe \SU{12} models with interesting
features that were constructed with the help of a Mathematica computer 
package written by one of the authors (RPF) called LieART \cite{Feger:2012bs}. 
This allows one to compute tensor products, branching rules, etc.,
and perform detailed searches for satisfactory models, although the predictions
of such models are limited by the number of parameters needed to describe the 
data. We find that after all the known quark and lepton mass and mixing data 
are used to fit the data to our models, some predictions arise for the 
yet-unknown results for the neutrino mass hierarchy and individual masses, 
leptonic CP violation, 
octant for the atmospheric mixing angle, and the effective mass that can be 
observed in neutrinoless double beta decay.

Expanding the gauge group to eliminate all or part of the family and flavor 
symmetries has been discussed previously, see references  \cite{Barr:2008gz,
Barr:2008pn,Dent:2009pd}. An earlier version of an \SU{12} model was 
previously published~\cite{Albright:2012zt,Albright:2012bp}, but 
subsequently several issues were found with some of the details, which are 
corrected here. In addition, we have adopted a new approach, made a more 
extensive study of the possibilities within this \SU{12} framework,  and 
present a more comprehensive treatment of these models, all of which are 
discussed below.

\section{INGREDIENTS OF A UNIFICATION GROUP}

Our starting point is a supersymmetric \SU{N} unification group, where $N$
must be large enough to assign chiral \SU{N} matter families to a number of 
irreps without the need for a flavor symmetry to distinguish the families.  
In practice this requires $N \geq 8$, while models derived from orbifold 
compactifications of SO(32) and the heterotic string suggest $N \leq 14$ 
\cite{Nilles:2006np}. The larger \SU{N} GUT group
replaces both the conventional GUT and the flavor groups cited earlier. 

A crucial issue then concerns the breaking of the large \SU{N} group to a 
smaller GUT/family group such as \SU{5} which we choose for the rest of this 
paper. We consider symmetry breaking that occurs in 
two possible ways. In the conventional approach, the symmetry is broken
one step at a time with the help of the \SU{N} adjoint scalar fields:
\begin{equation}\label{eq:rankbyrank}
  \SU{N} \rightarrow \SU{N-1} \times \U{1} \rightarrow ... \rightarrow 
        \SU{5} \times \U{1}^{N-5}.
\end{equation}

\noindent Then complex irreps are typically needed to break the \U{1}'s and 
reduce the rank to 4.  (This choice was improperly made in 
\cite{Albright:2012zt} and negates some of the 
results of that paper.)  The other choice which we employ here 
reduces the rank in one step without any \U{1}'s occurring, i.e., \SU{N} 
$\rightarrow$ \SU{5}.  This direct
breaking preserves SUSY provided $N$ is even, no \SU{N}
adjoint is present, and the F-flat and D-flat conditions hold.  As shown in 
\cite{Frampton:1981pf,Frampton:1982mj,Buccella:1982nx}, a dramatic reduction 
in rank is possible provided the sum of the Dynkin weights vanishes for the 
vacuum expectation values (VEVs) involved in lowering the rank. This  
possibility exists for $N = 12$, as is easily demonstrated in Appendix A.

Other necessary conditions for a satisfactory unification group are the 
following. The matter fields must form anomaly-free sets of \SU{N} and 
\SU{5} irreps with three \SU{5} families. Restrictions on the Higgs fields 
also must obtain. The \SU{5} Higgs singlets must arise from \SU{N} conjugate 
pairs to ensure D-flat directions, and they acquire \SU{5} VEVs at the \SU{5}
GUT scale where the separation of scales is given by 
$M_{\SU{5}}/M_{\SU{N}} \sim 1/50$. With the SUSY GUT scale 
occurring around $2 \times 10^{16}$ GeV, this implies the \SU{N} scale can be 
as high as $10^{18}$ GeV, very close to the string scale. 
In addition, an \SU{5} adjoint ${\bf 24}$ should be present
to break the \SU{5} symmetry to the SM, but this adjoint should not be 
contained in an \SU{N} adjoint which would spoil the desired symmetry-breaking
pattern. One set or mixtures of two sets of Higgs doublets in ${\bf 5}$ 
and $\overline{\bf 5}$ of \SU{5} must be available to break the electroweak 
symmetry at the weak scale. The addition of massive matter pairs at the 
\SU{N} scale will then allow one to introduce an effective operator 
approach.

\section{SU(12) UNIFICATION MODELS}
\label{sec:SU12UnificationModels}

After an extensive, but not exhaustive scan of possible \SU{N} models, we 
have found a relatively economical set of models for $N=12$.  Thus, for the 
rest of this paper we  confine our attention primarily to the \SU{12} 
unification group. This group has twelve antisymmetric irreps, ten of which 
are complex, while the ${\bf 924}$ and singlet are real:
\begin{equation}\label{eq:su12irreps}
\mathbf{12,\ 66,\ 220,\ 495,\ 792,\ (924),\ \overline{792},\ \overline{495},\ 
  \overline{220},\ \overline{66},\ \overline{12},\ (1)}
\end{equation}

\noindent which can be represented by Young diagrams with one to twelve blocks 
stacked vertically in a single column. These irreps contain no \SU{5} exotics. 
Among the smaller anomaly-free sets containing exactly three families of 
\SU{5} fermion matter are the following:
\begin{equation}
\label{eq:afsets}
\begin{array}{lr}
  {\bf 66} + {\bf 495} + 2({\bf \overline{220}}) + 2({\bf \overline{12}})& \\
  {\bf 220} + 3({\bf \overline{66}}) + 3({\bf \overline{12}})& \\
  3({\bf 12}) + {\bf 220} + {\bf 792} + {\bf \overline{495}} + 
    3({\bf \overline{66}})& \\
  {\bf 66} + 2({\bf 495}) + {\bf \overline{792}} + 
    2({\bf \overline{220}}) + 8({\bf \overline{12}})& \\
  {\bf 220} + {\bf 495} + {\bf \overline{792}} + 3({\bf \overline{66}})
    + 9({\bf \overline{12}})& \\
  3({\bf 66}) + 3({\bf 792}) + 3({\bf \overline{495}}) + 6({\bf \overline{12}})
    & \\
  3({\bf 66}) + 2({\bf 792}) + 2({\bf \overline{495}}) + 12({\bf \overline{12}})
    & \\
\end{array}
\end{equation}

\noindent where we assume any complex conjugate pairs of irreps become massive 
at the \SU{12} scale.  

Two of the anomaly-free sets, the first and fourth, are of special interest
for the third family top and bottom quarks are neatly 
contained in the ${\bf 66}$ which has one 10-dimensional irrep in the \SU{5} 
subgroup.   For the fourth set, the rank-7 \SU{12}/\SU{5} factor group can be 
completely broken in one step while preserving supersymmetry with the aid of 
the \SU{5}-singlet chiral superpartners of the fermions acquiring VEVs at the 
\SU{12} unification scale.  For the simplest first set, one needs the help of 
one additional scalar pair acquiring a VEV at the unification scale.  
Examples are illustrated in Appendix A.
\newpage
With the aid of the \SU{12} $\rightarrow$ \SU{5} branching rules:\\
\begin{equation}\label{eq:brrules}
\begin{array}{rcrrrrr}
{\bf 66} &\rightarrow &  7({\bf 5})\ + &  ({\bf 10})\ + \ &   &  + & 
  21({\bf 1}),\\
{\bf 495} &\rightarrow &  35({\bf 5})\ + &  21({\bf 10})\ + &  7(\overline{\bf 
  10})\ + & (\overline{\bf 5})\ + &  35({\bf 1}),\\
\overline{\bf 792} &\rightarrow &  7({\bf 5})\ + &  21({\bf 10})\ + &  
  35(\overline{\bf 10})\ + & 35(\overline{\bf 5})\ + & 22({\bf 1}),\\
\overline{\bf 220} &\rightarrow &  &  ({\bf 10})\ + &  7(\overline{\bf 10})\ 
  + & 21(\overline{\bf 5})\ + & 35({\bf 1}),\\
\overline{\bf 12} &\rightarrow &  &  &  & (\overline{\bf 5})\ + &  7({\bf 1}).\\
\end{array}
\end{equation}

\noindent one can see that the two anomaly-free sets of interest break at the
\SU{5} scale to the following sets of \SU{5} irreps according to 
\begin{equation} 
\begin{array}{rcl}
\label{eq:specset}
  {\bf 66}{+}{\bf 495}{+}2({\bf \overline{220}}){+}2({\bf \overline{12}}) 
  &\rightarrow& 3({\bf 10 + \overline{5} + 1}) + 21({\bf 10 + \overline{10}})
    + 42({\bf 5 + \overline{5}}) + 137({\bf 1}),\\
  {\bf 66}{+}2({\bf 495}){+}{\bf \overline{792}}{+}2({\bf \overline{220}}){+}8({\bf \overline{12}})
  &\rightarrow& 3({\bf 10 + \overline{5} + 1}) + 63({\bf 10 + \overline{10}}) 
    + 84({\bf 5 + \overline{5}}) + 236({\bf 1}),\\
\end{array}
\end{equation}

\noindent where both have three chiral families containing the observed 
lefthanded quarks and leptons and lefthanded antiquarks and antileptons.
The conjugate paired irreps all become massive at the SU(12) scale and are of
no more interest to us.

The three \SU{5} families can then be selected from among the following:
\begin{equation}
\label{eq:familychoices}
\begin{array}{lcrlcr}
({\bf 10}){\bf 66} &=& (2 + 0), \qquad&\qquad 
  ({\bf \overline{5}}){\bf 495} &=& (4 + 0),\\
({\bf 10}){\bf 495} &=& (2 + 2), \qquad&\qquad 
  ({\bf \overline{5}})\overline{\bf 792} &=& (4 + 3),\\
({\bf 10})\overline{\bf 792} &=& (2 + 5), \qquad&\qquad 
  ({\bf \overline{5}})\overline{\bf 220} &=& (4 + 5),\\
({\bf 10})\overline{\bf 220} &=& (2 + 7), \qquad&\qquad 
  ({\bf \overline{5}})\overline{\bf 12} &=& (4 + 7),\\
\end{array}
\end{equation}

\noindent where by Eq.~\eqref{eq:specset} up to two sets of ${\bf 495}$ and 
${\bf \overline{220}}$ and possibly more for ${\bf \overline{12}}$  are 
available for selection.  For our purposes, no discrete symmetry is needed 
to distinguish them.  We have chosen an \SU{12} basis $(a+b)$, where the first 
number $a$ in parenthesis refers to the number of \SU{5} boxes placed on top 
of the second remaining number $b$ of \SU{12}/\SU{5} boxes in the column of 
Young diagram boxes.  If two columns are present in a diagram representing 
higher dimensional irreps, the two pairs of numbers will be 
separated by a comma. 

Singlet Higgs conjugate pairs can be selected from among:
\begin{equation}
\label{eq:singlethiggs}
\begin{array}{rclrcl}
({\bf 1})\higgs{12} &=& (0 + 1), \qquad&\qquad ({\bf 1})\higgsbar{12} 
  &=& (5 + 6),\\
({\bf 1})\higgs{66} &=& (0 + 2), \qquad&\qquad ({\bf 1})\higgsbar{66} 
  &=& (5 + 5),\\
({\bf 1})\higgs{220} &=& (0 + 3), \qquad&\qquad ({\bf 1})\higgsbar{220}
  &=& (5 + 4),\\
({\bf 1})\higgs{495} &=& (0 + 4), \qquad&\qquad ({\bf 1})\higgsbar{495}
  &=& (5 + 3),\\
({\bf 1})\higgs{792} &=& (0 + 5)\ {\rm or}\ (5 + 0), \qquad&\qquad 
  ({\bf 1})\higgsbar{792} &=& (5 + 2)\ {\rm or}\ (0 + 7).\\
\end{array}
\end{equation}

\noindent  For simplicity we shall assume that the VEVs of the SU(5) Higgs 
singlets chosen in each model and their couplings to fermions are real and 
equal.

As emphasized earlier, a 24-plet Higgs, which must be present to break the 
\SU{5} GUT symmetry down to the SM, can not be part of the \SU{12} adjoint 
${\bf 143}$ in the one-step breaking of \SU{12} to \SU{5}.  Instead, we find 
it best to include the \SU{5} adjoint in the complex pair of 
$({\bf 24})\higgs{5148}$
and $({\bf 24})\higgsbar{5148}$ Higgs irreps which can develop VEVs at 
the GUT scale.  In fact, the SU(12) breaking of this $\bf{5148}$ and  
$\bf{\overline{5148}}$ pair yields only one $({\bf 24})$ each, as can be seen 
from the following decomposition for the $\bf{5148}$,  
\begin{equation}
\label{eq:adjointhiggs}
{\bf 5148} \rightarrow ({\bf 24}) +  245({\bf 5}) + 147({\bf 10}) +
  49({\bf \overline{10}}) + 7({\bf \overline{5}}) + 35({\bf 15}) + 
  21({\bf \overline{40}}) + 7({\bf \overline{45}}) + 224({\bf 1}),
\end{equation}

\noindent and similarly for the conjugate irrep. The (24) Higgs contributions 
are represented by 
\begin{equation}
\begin{array}{rclrcl}
  ({\bf 24})\higgs{5148} &=& (4 + 0, 1 + 0),\qquad&\qquad 
    ({\bf 24})\higgsbar{5148} 
    &=& (4 + 7, 1 + 7).\\
\end{array}
\end{equation}

\noindent   Because these irreps represent complex pairs, we shall also 
assume that their VEVs are complex conjugates of each other and assign a 
common VEV to the quarks and a different common VEV to the leptons.  This 
can be accomplished if their VEVs point in the $2B-L$ direction which is a 
linear combination of the $\lambda_{15}$ and $\lambda_{24}$ generators of SU(5): 
\begin{equation}
\begin{array}{rcl}
  2B - L &=& (5/12)\sqrt{6}\lambda_{15} + (1/4)\sqrt{10}\lambda_{24}\\
         &=& {\rm diag}(2/3,\ 2/3,\ 2/3,\ -1,\ -1)
\end{array}
\end{equation}

\noindent We then adopt the following notation for their VEVs:
\begin{equation}\label{eqn:5148VEVs}
  \langle ({\bf 24})\higgs{5148}\rangle = (2B-L)\kappa,\qquad 
     \langle ({\bf 24})\higgsbar{5148}\rangle = (2B-L)\kappa^*,
\end{equation}

\noindent  Hence this choice provides 
a ready way in which to introduce complex phases into the mass matrices. The 
different VEVs generated from these Higgs fields will also prove useful to 
break the down-quark and charged-lepton mass spectral degeneracy.

In addition, we need a Higgs singlet to give mass to the lefthanded conjugate 
neutrinos.  Since all families of such neutrinos are in SU(5) and SU(12) 
singlets, it is convenient to introduce a $({\bf 1}){\bf 1_H}$ Higgs singlet 
for this purpose.  A dim-4 vertex mass diagram then requires that this Higgs 
singlet must change lepton number by two units, or $\Delta L = + 2$.

In general, two sets of Higgs doublets which remain light down to the 
EW scale where they get VEVs can be formed from linear combinations of the
${\bf 5}$'s and ${\bf \overline{5}}$'s of \SU{5}: 
\begin{equation}
\label{eq:EWhiggs}
a_1 ({\bf 5})\higgs{12} + a_2 ({\bf 5})\higgsbar{495},\quad 
  {\rm and}\quad   
b_1 ({\bf \overline{5}})\higgsbar{12} + b_2 
  ({\bf \overline{5}})\higgs{495}.
\end{equation}

\noindent In what follows in Sect. IV., it will become apparent that the 
$({\bf 5})\higgsbar{495}$ must develop an EW VEV, while the 
$({\bf 5})\higgs{12}$
can get massive without requiring that it also develops an EW VEV.  The 
situation is not so clear-cut for an EW $({\bf \overline{5}})$ VEV generated 
from the $\higgs{495}$ or $\higgsbar{12}$ Higgs.  We shall consider both
cases individually in our search for models and comment later on the results.

Renormalizable dim-4 operators can be formed from three-point vertices involving
two fermions and a Higgs. This requires we identify the appropriate \SU{5} and
\SU{12} singlet vertices. For this purpose, Young diagram product rules  
must be applied at every vertex, so that the \SU{5} boxes are on top of the
remaining \SU{12}/\SU{5} boxes.  For example, 
$({\bf 10}){\bf \overline{220}} . ({\bf \overline{5}}){\bf \overline{12}} . 
({\bf \overline{5}})\higgs{495} = (2 + 7) . (4 + 7) . (4 + 0) = 
(2 + 7) . (5 + 7, 3 + 0) = (5 + 7, 5 + 7)$
is a proper \SU{5}- and \SU{12}-singlet vertex with two columns of 
12 boxes with the 5 \SU{5} boxes on top; on the other hand, the product 
$({\bf 10}){\bf 792} . ({\bf \overline{5}}){\bf \overline{495}} . 
({\bf \overline{5}})\higgsbar{12} = (2 + 3) . (4 + 4) . (4 + 7)$ 
is not, 
for one can not carry out the product keeping the 5 \SU{5} boxes on top of 
the remaining 7 \SU{12} boxes in both columns without rearrangements. 

Effective higher-dimensional operators can be formed by inserting  
Higgs and massive fermions in tree diagrams.  With SUSY valid at the \SU{5}
scale, loop diagrams are highly suppressed.  The massive intermediate fermion
pairs at the \SU{12} scale, which are formed from complex irreps and are 
obviously anomaly-free, can 
be selected from among the
\begin{equation}
\begin{array}{rl}
  & {\bf 12} \times \bf{\overline{12}},\ \bf{66} \times \bf{\overline{66}},
    \ \bf{220} \times \bf{\overline{220}},\ \bf{495} \times 
    \bf{\overline{495}},\ \bf{792} \times \bf{\overline{792}}\\
\end{array}
\end{equation}

\noindent pair insertions.  
In order to maintain the proper basis with the \SU{5} boxes at the top of 
the Young diagrams, the only proper contractions of interest here involve 
the following:
\begin{equation}
\label{eq:massivefermions}
\begin{array}{rclrcl}
  ({\bf 1}){\bf 12} &\times& ({\bf 1}){\bf \overline{12}},&\qquad 
    ({\bf 5}){\bf 12} &\times& ({\bf \overline{5}}){\bf \overline{12}},\\
  ({\bf 1}){\bf 66} &\times& ({\bf 1}){\bf \overline{66}},&\qquad 
    ({\bf 10}){\bf 66} &\times& ({\bf \overline{10}}){\bf \overline{66}},\\
  ({\bf 1}){\bf 220} &\times& ({\bf 1}){\bf \overline{220}},&\qquad 
      ({\bf \overline{10}}){\bf 220} &\times& ({\bf 10}){\bf \overline{220}},\\
  ({\bf 1}){\bf 495} &\times& ({\bf 1}){\bf \overline{495}},&\qquad 
      ({\bf \overline{5}}){\bf 495} &\times& ({\bf 5}){\bf \overline{495}},\\
  ({\bf 1}){\bf 792} &\times& ({\bf 1}){\bf \overline{792}}.\\
\end{array}
\end{equation}

We can now proceed to construct the most general Higgs and Yukawa 
superpotentials preserving R-parity, where the Higgs superfields and the 
matter superfields are assigned R-parity +1.  The Higgs superpotential with 
the three-point couplings involving all Higgs fields 
which appear in \SU{12} has the following \SU{5} and \SU{12} singlet terms:
\begin{equation}
\label{eq:higgsW}
\begin{array}{rcl}
  W_{\rm Higgs} &=& ({\bf 1})\higgs{12} . ({\bf 1})\higgsbar{12} 
    + ({\bf 1})\higgs{66} . ({\bf 1})\higgsbar{66} 
    + ({\bf 1})\higgs{220} . ({\bf 1})\higgsbar{220} 
    + ({\bf 1})\higgs{495} . ({\bf 1})\higgsbar{495}\\ 
   &+& ({\bf 1})\higgs{792} . ({\bf 1})\higgsbar{792}
    + ({\bf 24})\higgs{5148} . ({\bf 24})\higgsbar{5148}
    + ({\bf \overline{5}})\higgsbar{12} . ({\bf 5})\higgsbar{495} . 
      ({\bf 24})\higgs{5148}\\
   &+& ({\bf \overline{5}})\higgs{495} . ({\bf 5})\higgs{12} . ({\bf 24})\higgsbar{5148}
    + ({\bf \overline{5}})\higgsbar{12} . ({\bf 5})\higgsbar{495} .
      ({\bf 1})\higgs{792}\\
   &+& ({\bf \overline{5}})\higgs{495} . ({\bf 5})\higgs{12} . ({\bf 1})\higgsbar{792}
    + ({\bf 1})\higgs{12} . ({\bf 1})\higgs{12} . ({\bf 1})\higgsbar{66} 
    + ({\bf 1})\higgs{12} . ({\bf 1})\higgs{66} . ({\bf 1})\higgsbar{220}\\
   &+& ({\bf 1})\higgs{12} . ({\bf 1})\higgs{220} . ({\bf 1})\higgsbar{495}
    + ({\bf 1})\higgs{12} . ({\bf 1})\higgs{495} . ({\bf 1})\higgsbar{792}
    + ({\bf 1})\higgs{66} . ({\bf 1})\higgs{66} . ({\bf 1})\higgsbar{495}\\
   &+&({\bf 1})\higgs{66} . ({\bf 1})\higgs{220} . ({\bf 1})\higgsbar{792}
    +({\bf 1})\higgs{66} . ({\bf 1})\higgs{792} . ({\bf 1})\higgs{792}
    + ({\bf 1})\higgs{66} . ({\bf 1})\higgsbar{12} . ({\bf 1})\higgsbar{12}\\
   &+& ({\bf 1})\higgs{220} . ({\bf 1})\higgs{495} . ({\bf 1})\higgs{792} 
    + ({\bf 1})\higgs{220} . ({\bf 1})\higgsbar{66} . ({\bf 1})\higgsbar{12}
    + ({\bf 1})\higgs{495} . ({\bf 1})\higgsbar{220} . ({\bf 1})\higgsbar{12}\\
   &+& ({\bf 1})\higgs{495} . ({\bf 1})\higgsbar{66} . ({\bf 1})\higgsbar{66}
    + ({\bf 1})\higgs{792} . ({\bf 1})\higgsbar{495} . ({\bf 1})\higgsbar{12}
    + ({\bf 1})\higgs{792} . ({\bf 1})\higgsbar{220} . ({\bf 1})\higgsbar{66}\\
   &+& ({\bf 1})\higgsbar{792} . ({\bf 1})\higgsbar{792} . 
      ({\bf 1})\higgsbar{66}
    + ({\bf 1})\higgsbar{792} . ({\bf 1})\higgsbar{495} . 
      ({\bf 1})\higgsbar{220}\\
\end{array}
\end{equation}

The corresponding Yukawa superpotential has the following structure:
\begin{equation}
\label{eq:yukawaw}
W_{\rm Yukawa} = W_{({\bf 24})} + W_{({\bf 5})} + W_{({\bf \overline{5}})} 
   + W_{({\bf 1})},\\
\end{equation}
where
\begin{displaymath}
\begin{array}{rcl}
W_{({\bf 24})} &=& ({\bf \overline{10}}){\bf \overline{66}} . ({\bf 10}){\bf \overline{220}} .
     ({\bf 24})\higgs{5148}
   + ({\bf \overline{10}}){\bf 220} . ({\bf 10}){\bf 66} . ({\bf 24})\higgsbar{5148}\\
  &+& ({\bf \overline{5}}){\bf \overline{12}} . ({\bf 5}){\bf \overline{495}} .
     ({\bf 24})\higgs{5148}
   + ({\bf \overline{5}}){\bf 495} . ({\bf 5}){\bf 12} . ({\bf 24})\higgsbar{5148},
   \\[0.2in]
W_{({\bf 5})} &=& ({\bf 10}){\bf \overline{220}} . ({\bf 10}){\bf 66} . ({\bf 5})\higgs{12}
    + ({\bf \overline{10}}){\bf 220} . ({\bf 5}){\bf \overline{495}} . ({\bf 5})\higgs{12}
    + ({\bf \overline{10}}){\bf \overline{66}} . ({\bf 5}){\bf 12} . ({\bf 5})\higgs{12}\\
   &+& ({\bf \overline{5}}){\bf \overline{66}} . ({\bf 1}){\bf 12} . ({\bf 5})\higgs{12}
    + ({\bf \overline{5}}){\bf \overline{220}} . ({\bf 1}){\bf 66} . ({\bf 5})\higgs{12}
    + ({\bf \overline{5}}){\bf \overline{495}} . ({\bf 1}){\bf 220} . ({\bf 5})\higgs{12}\\
   &+& ({\bf \overline{5}}){\bf \overline{792}} . ({\bf 1}){\bf 495} . ({\bf 5})\higgs{12}
    + ({\bf \overline{5}}){\bf 495} . ({\bf 1}){\bf \overline{792}} . ({\bf 5})\higgs{12}
    +  ({\bf 10}){\bf 66} . ({\bf 10}){\bf 66} . ({\bf 5})\higgsbar{495}\\
   &+& ({\bf \overline{10}}){\bf 220} . ({\bf 5}){\bf 12} . ({\bf 5})\higgsbar{495}
    + ({\bf \overline{5}}){\bf \overline{12}} . ({\bf 1}){\bf 792} .
      ({\bf 5})\higgsbar{495}
    + ({\bf \overline{5}}){\bf \overline{220}} . ({\bf 1}){\bf \overline{792}} . 
      ({\bf 5})\higgsbar{495}\\
   &+& ({\bf \overline{5}}){\bf \overline{495}} . ({\bf 1}){\bf \overline{495}} . 
      ({\bf 5})\higgsbar{495}
    + ({\bf \overline{5}}){\bf \overline{792}} . ({\bf 1}){\bf \overline{220}} .
      ({\bf 5})\higgsbar{495}\\
   &+& ({\bf \overline{5}}){\bf 792} . ({\bf 1}){\bf \overline{12}} . 
      ({\bf 5})\higgsbar{495},\\[0.2in]
W_{({\bf \overline{5}})} &=& ({\bf \overline{10}}){\bf \overline{66}} . 
      ({\bf \overline{10}}){\bf \overline{66}} . ({\bf \overline{5}})\higgs{495}
    + ({\bf 10}){\bf \overline{220}} . ({\bf \overline{5}}){\bf \overline{12}} . 
      ({\bf \overline{5}})\higgs{495}
    + ({\bf 5}){\bf \overline{792}} . ({\bf 1}){\bf 12} . ({\bf \overline{5}})\higgs{495}\\
   &+& ({\bf 5}){\bf 792} . ({\bf 1}){\bf 220} . (\bf \overline{5})\higgs{495}
    + ({\bf 5}){\bf 495} . ({\bf 1}){\bf 495} . ({\bf \overline{5}})\higgs{495}
    + ({\bf 5}){\bf 220} . ({\bf 1}){\bf 792} . ({\bf \overline{5}})\higgs{495}\\
   &+& ({\bf 5}){\bf 12} . ({\bf 1}){\bf \overline{792}} . ({\bf \overline{5}})\higgs{495}
    + ({\bf \overline{10}}){\bf \overline{66}} . 
      ({\bf \overline{10}}){\bf 220} . ({\bf \overline{5}})\higgsbar{12}
    + ({\bf 10}){\bf \overline{220}} . ({\bf \overline{5}}){\bf 495} . 
      ({\bf \overline{5}})\higgsbar{12}\\
   &+& ({\bf 10}){\bf 66} . ({\bf \overline{5}}){\bf \overline{12}} .
      ({\bf \overline{5}})\higgsbar{12}
    + ({\bf 5}){\bf \overline{495}} . ({\bf 1}){\bf 792} . 
      ({\bf \overline{5}})\higgsbar{12}
    + ({\bf 5}){\bf 792} . ({\bf 1}){\bf \overline{495}} . 
      ({\bf \overline{5}})\higgsbar{12}\\
   &+& ({\bf 5}){\bf 495} . ({\bf 1}){\bf \overline{220}} . 
      ({\bf \overline{5}})\higgsbar{12}
    + ({\bf 5}){\bf 220} . ({\bf 1}){\bf \overline{66}} . 
      ({\bf \overline{5}})\higgsbar{12}
    + ({\bf 5}){\bf 66} . ({\bf 1}){\bf \overline{12}} . 
      ({\bf \overline{5}})\higgsbar{12},\\
\end{array}
\end{displaymath}
\newpage
\begin{displaymath}
\begin{array}{rcl}
W_{({\bf 1})} &=& ({\bf \overline{10}}){\bf 220} . ({\bf 10}){\bf \overline{495}} . ({\bf 1})\higgs{12}
    + ({\bf \overline{10}}){\bf \overline{220}} . ({\bf 10}){\bf 66} . ({\bf 1})\higgs{12}
    + ({\bf \overline{10}}){\bf 220} . ({\bf 10}){\bf \overline{792}} . ({\bf 1})\higgs{66}\\
   &+& ({\bf \overline{10}}){\bf \overline{495}} . ({\bf 10}){\bf 66} . ({\bf 1})\higgs{66}
    + ({\bf \overline{10}}){\bf \overline{792}} . ({\bf 10}){\bf 66} . ({\bf 1})\higgs{220}
    + ({\bf \overline{10}}){\bf 220} . ({\bf 10}){\bf 792} . ({\bf 1})\higgs{495}\\
   &+& ({\bf \overline{10}}){\bf 220} . ({\bf 10}){\bf 495} . ({\bf 1})\higgs{792}
    + ({\bf \overline{10}}){\bf 792} . ({\bf 10}){\bf 66} . ({\bf 1})\higgs{792}
    + ({\bf \overline{10}}){\bf \overline{66}} . ({\bf 10}){\bf \overline{220}} .
      ({\bf 1})\higgs{792}\\
   &+& ({\bf \overline{10}}){\bf 220} . ({\bf 10}){\bf 66} . ({\bf 1})\higgsbar{792}
    + ({\bf \overline{10}}){\bf \overline{495}} . ({\bf 10}){\bf \overline{220}} .
      ({\bf 1})\higgsbar{792}
    + ({\bf \overline{10}}){\bf \overline{66}} . ({\bf 10}){\bf \overline{792}} . 
      ({\bf 1})\higgsbar{792}\\
   &+& ({\bf \overline{10}}){\bf \overline{792}} . ({\bf 10}){\bf \overline{220}} . 
      ({\bf 1})\higgsbar{495}
    + ({\bf \overline{10}}){\bf \overline{66}} . ({\bf 10}){\bf 792} . 
      ({\bf 1})\higgsbar{220}
    + ({\bf \overline{10}}){\bf 792} . ({\bf 10}){\bf \overline{220}} . 
      ({\bf 1})\higgsbar{66}\\
   &+& ({\bf \overline{10}}){\bf \overline{66}} . ({\bf 10}){\bf 495} . 
      ({\bf 1})\higgsbar{66}
    + ({\bf \overline{10}}){\bf 495} . ({\bf 10}){\bf \overline{220}} . 
      ({\bf 1})\higgsbar{12}
    + ({\bf \overline{10}}){\bf \overline{66}} . ({\bf 10}){\bf 220} . 
      ({\bf 1})\higgsbar{12}\\
   &+& ({\bf \overline{5}}){\bf 495} . ({\bf 5}){\bf \overline{792}} . ({\bf 1})\higgs{12}
    + ({\bf \overline{5}}){\bf \overline{66}} . ({\bf 5}){\bf 12} . ({\bf 1})\higgs{12}
    + ({\bf \overline{5}}){\bf \overline{220}} . ({\bf 5}){\bf 12} . ({\bf 1})\higgs{66}\\
   &+& ({\bf \overline{5}}){\bf 495} . ({\bf 5}){\bf 792} . ({\bf 1})\higgs{220}
    + ({\bf \overline{5}}){\bf \overline{495}} . ({\bf 5}){\bf 12} . ({\bf 1})\higgs{220}
    + ({\bf \overline{5}}){\bf 495} . ({\bf 5}){\bf 495} . ({\bf 1})\higgs{495}\\
   &+& ({\bf \overline{5}}){\bf \overline{792}} . ({\bf 5}){\bf 12} . ({\bf 1})\higgs{495}
    + ({\bf \overline{5}}){\bf 495} . ({\bf 5}){\bf 220} . ({\bf 1})\higgs{792}
    + ({\bf \overline{5}}){\bf \overline{12}} . ({\bf 5}){\bf \overline{495}} . 
      ({\bf 1})\higgs{792}\\
   &+& ({\bf \overline{5}}){\bf 495} . ({\bf 5}){\bf 12} . ({\bf 1})\higgsbar{792}
    + ({\bf \overline{5}}){\bf \overline{220}} . ({\bf 5}){\bf \overline{495}} .
      ({\bf 1})\higgsbar{792}
    + ({\bf \overline{5}}){\bf \overline{495}} . ({\bf 5}){\bf \overline{495}} . 
      ({\bf 1})\higgsbar{495}\\
   &+& ({\bf \overline{5}}){\bf \overline{12}} . ({\bf 5}){\bf 792} . 
      ({\bf 1})\higgsbar{495}
    + ({\bf \overline{5}}){\bf \overline{792}} . ({\bf 5}){\bf \overline{495}} .
      ({\bf 1})\higgsbar{220}
    + ({\bf \overline{5}}){\bf \overline{12}} . ({\bf 5}){\bf 495} . 
      ({\bf 1})\higgsbar{220}\\
   &+& ({\bf \overline{5}}){\bf \overline{12}} . ({\bf 5}){\bf 220} . 
      ({\bf 1})\higgsbar{66}
    + ({\bf \overline{5}}){\bf 792} . ({\bf 5}){\bf \overline{495}} .
      ({\bf 1})\higgsbar{12}
    + ({\bf \overline{5}}){\bf \overline{12}} . ({\bf 5}){\bf 66} . 
      ({\bf 1})\higgsbar{12}\\
  &+& ({\bf 1})\higgs{12} . ({\bf 1}){\bf 12} . ({\bf 1}){\bf \overline{66}} 
    + ({\bf 1})\higgs{12} . ({\bf 1}){\bf 66} . ({\bf 1}){\bf \overline{220}}
    + ({\bf 1})\higgs{12} . ({\bf 1}){\bf 220} . ({\bf 1}){\bf \overline{495}}\\ 
   &+& ({\bf 1})\higgs{12} . ({\bf 1}){\bf 495} . ({\bf 1}){\bf \overline{792}}
    + ({\bf 1})\higgs{66} . ({\bf 1}){\bf 66} . ({\bf 1}){\bf \overline{495}}
    + ({\bf 1})\higgs{66} . ({\bf 1}){\bf 220} . ({\bf 1}){\bf \overline{792}}\\  
   &+& ({\bf 1})\higgs{66} . ({\bf 1}){\bf 792} . ({\bf 1}){\bf 792}
    + ({\bf 1})\higgs{66} . ({\bf 1}){\bf \overline{12}} . ({\bf 1}){\bf \overline{12}}
    + ({\bf 1})\higgs{220} . ({\bf 1}){\bf 495} . ({\bf 1}){\bf 792}\\
   &+& ({\bf 1})\higgs{220} . ({\bf 1}){\bf \overline{66}} . ({\bf 1}){\bf \overline{12}}
    + ({\bf 1})\higgs{495} . ({\bf 1}){\bf \overline{220}} . ({\bf 1}){\bf \overline{12}}
    + ({\bf 1})\higgs{495} . ({\bf 1}){\bf \overline{66}} . ({\bf 1}){\bf \overline{66}}\\
   &+& ({\bf 1})\higgs{792} . ({\bf 1}){\bf \overline{495}} . ({\bf 1}){\bf \overline{12}}
    + ({\bf 1})\higgs{792} . ({\bf 1}){\bf \overline{220}} . ({\bf 1}){\bf \overline{66}}
    + ({\bf 1})\higgsbar{792} . ({\bf 1}){\bf \overline{792}} . 
      ({\bf 1}){\bf \overline{66}}\\
   &+& ({\bf 1})\higgsbar{792} . ({\bf 1}){\bf \overline{495}} . 
      ({\bf 1}){\bf \overline{220}}
    + ({\bf 1}){\bf 12} . ({\bf 1})\higgs{12} . ({\bf 1}){\bf \overline{66}} 
    + ({\bf 1}){\bf 12} . ({\bf 1})\higgs{66} . ({\bf 1}){\bf \overline{220}}\\
   &+& ({\bf 1}){\bf 12} . ({\bf 1})\higgs{220} . ({\bf 1}){\bf \overline{495}} 
    + ({\bf 1}){\bf 12} . ({\bf 1})\higgs{495} . ({\bf 1}){\bf \overline{792}}
    + ({\bf 1}){\bf 66} . ({\bf 1})\higgs{66} . ({\bf 1}){\bf \overline{495}}\\
   &+& ({\bf 1}){\bf 66} . ({\bf 1})\higgs{220} . ({\bf 1}){\bf \overline{792}}  
    + ({\bf 1}){\bf 66} . ({\bf 1})\higgs{792} . ({\bf 1}){\bf 792}
    + ({\bf 1}){\bf 66} . ({\bf 1})\higgsbar{12} . ({\bf 1}){\bf \overline{12}}\\
   &+& ({\bf 1}){\bf 220} . ({\bf 1})\higgs{495} . ({\bf 1}){\bf 792} 
    + ({\bf 1}){\bf 220} . ({\bf 1})\higgsbar{66} . ({\bf 1}){\bf \overline{12}}
    + ({\bf 1}){\bf 495} . ({\bf 1})\higgsbar{220} . ({\bf 1}){\bf \overline{12}}\\
   &+& ({\bf 1}){\bf 495} . ({\bf 1})\higgsbar{66} . ({\bf 1}){\bf \overline{66}}
    + ({\bf 1}){\bf 792} . ({\bf 1})\higgsbar{495} . ({\bf 1}){\bf \overline{12}}
    + ({\bf 1}){\bf 792} . ({\bf 1})\higgsbar{220} . ({\bf 1}){\bf \overline{66}}\\
   &+& ({\bf 1}){\bf \overline{792}} . ({\bf 1})\higgsbar{792} . 
      ({\bf 1}){\bf \overline{66}} 
    + ({\bf 1}){\bf \overline{792}} . ({\bf 1})\higgsbar{495} . 
      ({\bf 1}){\bf \overline{220}}
    + ({\bf 1}){\bf 12} . ({\bf 1}){\bf 12} . ({\bf 1})\higgsbar{66}\\
   &+& ({\bf 1}){\bf 12} . ({\bf 1}){\bf 66} . ({\bf 1})\higgsbar{220}
    + ({\bf 1}){\bf 12} . ({\bf 1}){\bf 220} . ({\bf 1})\higgsbar{495} 
    + ({\bf 1}){\bf 12} . ({\bf 1}){\bf 495} . ({\bf 1})\higgsbar{792}\\
   &+& ({\bf 1}){\bf 66} . ({\bf 1}){\bf 66} . ({\bf 1})\higgsbar{495}
    + ({\bf 1}){\bf 66} . ({\bf 1}){\bf 220} . ({\bf 1})\higgsbar{792}  
    + ({\bf 1}){\bf 66} . ({\bf 1}){\bf 792} . ({\bf 1})\higgs{792}\\
   &+& ({\bf 1}){\bf 66} . ({\bf 1}){\bf \overline{12}} . ({\bf 1})\higgsbar{12}
    + ({\bf 1}){\bf 220} . ({\bf 1}){\bf 495} . ({\bf 1})\higgs{792} 
    + ({\bf 1}){\bf 220} . ({\bf 1}){\bf \overline{66}} . ({\bf 1})\higgsbar{12}\\
   &+& ({\bf 1}){\bf 495} . ({\bf 1}){\bf \overline{220}} . ({\bf 1})\higgsbar{12}
    + ({\bf 1}){\bf 495} . ({\bf 1}){\bf \overline{66}} . ({\bf 1})\higgsbar{66}
    + ({\bf 1}){\bf 792} . ({\bf 1}){\bf \overline{495}} . ({\bf 1})\higgsbar{12}\\
   &+& ({\bf 1}){\bf 792} . ({\bf 1}){\bf \overline{220}} . ({\bf 1})\higgsbar{66}
    + ({\bf 1}){\bf \overline{792}} . ({\bf 1}){\bf \overline{792}} . 
      ({\bf 1})\higgsbar{66} 
    + ({\bf 1}){\bf \overline{792}} . ({\bf 1}){\bf \overline{495}} . 
      ({\bf 1})\higgsbar{220}.\\
\end{array}
\end{displaymath}

With these ingredients in mind, we can now construct \SU{12} models whose 
renormalizable and effective higher-dimensional operators determine the 
elements of the quark and lepton mass matrices.  The fitting procedure
to be described later then allows us to determine which models are viable and
acceptable in describing the quark and lepton mass and mixing data.

\section{SU(12) MODEL CONSTRUCTION WITH EFFECTIVE OPERATORS}
	
Starting with either the first or fourth anomaly-free sets of 
Eq.~\eqref{eq:specset}, we can assign \SU{12} irreps for the three \SU{5} 
$({\bf 10})$ family members defining the up quark mass matrix ($M_{\rm U}$), 
the three $({\bf \overline{5}})$ family members required in addition to define 
the down quark mass matrix ($M_{\rm D}$), and the additional 
three singlets defining the Dirac neutrino ($M_{\rm DN}$) and Majorana 
neutrino ($M_{\rm MN}$) mass matrices.  Because of the greater arbitrariness
in making these family assignments for the fourth anomaly-free set, we shall
concentrate our attention from now on to the simplest first anomaly-free set of
Eq.~(5). The contributions to the matrix
elements for the Yukawa matrices involving the up quarks (${\bf Uij}$), 
down quarks (${\bf Dij}$), charged leptons (${\bf Lij}$), and Dirac neutrinos 
(${\bf DNij}$), as well as the Majorana matrix for the heavy righthanded 
neutrinos (${\bf MNij}$), can arise from renormalizable dim-4 operators as well 
as higher dimensional effective operators involving \SU{5} ({\bf 1}) scalar 
singlets and ({\bf 24}) scalar adjoints appearing in external lines, along 
with a ({\bf 5}) or $({\bf \overline{5}})$ EW Higgs scalar in the case of the 
Yukawa matrices.  Each effective operator diagram must be constructed according
to the Young diagram multiplication rules illustrated in the previous section, 
where each vertex of the diagram represents a term in the superpotential of 
Eq.~\eqref{eq:yukawaw}.

\subsection{Possible Sets of Assignments for the Chiral Fermion Families}

We begin with the 33 component of the up quark Yukawa matrix (${\bf U33}$)
and strive for a dim-4 renormalizable contribution, as that will represent the 
largest source for the top quark mass.  Scanning through the four possible 
(${\bf 10}$) matter families and ({\bf 5}) Higgs assignments in 
Eqs.~\eqref{eq:familychoices} and \eqref{eq:EWhiggs}, it becomes clear that 
only one possibility exists for a proper 3-point vertex singlet, namely,
\begin{equation}\label{eq:U33}
\begin{array}{rl}
{\bf U33}:\quad & ({\bf 10}){\bf 66_3} . ({\bf 5})\higgsbar{495} . 
                           ({\bf 10}){\bf 66_3}\\
	& = (2 + 0) . (1 + 7) . (2 + 0) = (5 + 7) \sim ({\bf 1}){\bf 1}.\\
\end{array}
\end{equation}

\noindent  For all other ${\bf U}$ matrix elements, the effective operators 
will be dim-5 or higher, with one or more singlet and adjoint Higgs fields, 
as well as the $({\bf 5})\higgsbar{495}$ Higgs field attached to the fermion 
line.  We shall assume the other ({\bf 5})\irrepsub{12}{H} is inert and does
not develop a VEV.
  
For the simplest anomaly-free set of Eqs.~\eqref{eq:afsets} and 
\eqref{eq:specset}, the possible assignments of the ({\bf 10}) family members 
are 
$({\bf 10}){\bf 495},\ ({\bf 10}){\bf \overline{220}},\ ({\bf 10}){\bf 66}_3$ 
and its permutation of the first and second family assignments, along with  
$({\bf 10}){\bf \overline{220}},\ ({\bf 10}){\bf \overline{220}},\ ({\bf 10}){\bf 66}_3$.  The three $({\bf \overline{5}})$ family members can then be 
selected from the four possibilities given in Eq.~\eqref{eq:familychoices}, 
consistent with the anomaly-free set in question in Eq.~\eqref{eq:specset}.
We list below the permissible ({\bf 10}) and $({\bf \overline{5}})$ family combinations, 
\begin{equation}
\label{eq:familyids}
\begin{array}{ll}
  ({\bf 10}){\bf 495},\  ({\bf 10}){\bf \overline{220}},\ ({\bf 10}){\bf 66_3};
    \qquad\qquad & ({\bf \overline{5}}){\bf \overline{220}},\ 
    ({\bf \overline{5}}){\bf \overline{12}},\ ({\bf \overline{5}}){\bf \overline{12}};\\
  ({\bf 10}){\bf \overline{220}},\ ({\bf 10}){\bf \overline{220}},\ ({\bf 10}){\bf 66_3};
  \qquad\qquad & ({\bf \overline{5}}){\bf 495},\ ({\bf \overline{5}}){\bf \overline{12}},\ ({\bf \overline{5}}){\bf \overline{12}},\\
\end{array}
\end{equation}

\noindent where only one \SU{5} family is assigned to each of the \SU{12} 
irreps in the set.  It is to be understood that aside from the third ({\bf 10})
family member being associated with $({\bf 10}){\bf 66_3}$, all permutations of
the family assignments are allowed.  

Since all the non-trivial \SU{12} irreps in the set have already been assigned,
the conjugate lefthanded (or heavy righthanded) neutrinos must appear in stand 
alone \SU{12} singlet irreps, i.e., $({\bf 1}){\bf 1}$'s, one for each massive 
Majorana family: $({\bf 1}){\bf 1_1},\ ({\bf 1}){\bf 1_2},\ ({\bf 1}){\bf 1_3}$ 
with the assumption of three families of righthanded singlet neutrinos.  

\subsection{Construction of the Mass Matrix Elements}
\label{ssec:MassMatrixElementConstruction}

We now have all the necessary ingredients to assemble the renormalizable and 
effective operator contributions to the four Dirac and one Majorana mass
matrices.  We begin the actual mass matrix constructions with the 
${\bf U}$ matrix where, as noted earlier, the only suitable dim-4 contribution 
arises for the 33 element which involves the $({\bf 5})\higgsbar{495}$ 
EW Higgs, which we repeat here, 
\begin{equation}
{\bf U33:}\ ({\bf 10}){\bf 66_3} . ({\bf 5})\higgsbar{495} . ({\bf 10}){\bf 66_3}.
\end{equation}

In order to obtain an appropriate hierarchy for the ${\bf Uij}$
mass matrix elements, all other matrix elements must arise from dim-5 or higher
contributions involving the $({\bf 5})\higgsbar{495}$ and at least one  
singlet or adjoint Higgs field, and one or more massive fermion pairs. From 
the structure of the $M_{\rm U}$ matrix which involves only light 
({\bf 10}) \SU{5} chiral fermion families, it is clear that only ({\bf 10}) 
and $({\bf \overline{10}})$ massive fermions can contribute in the 
intermediate states.  From Eq.~\eqref{eq:massivefermions} it is then obvious 
that the only possible mass insertions will involve 
$({\bf 10}){\bf 66} \times ({\bf \overline{10}}){\bf \overline{66}}$ and/or  
$({\bf 10}){\bf \overline{220}} \times ({\bf \overline{10}}){\bf 220}$ irreps. 
As for the Higgs singlet vertices or those involving a Higgs ({\bf 24}), these 
will be determined by the list of Higgs fields considered, the light fermion 
families involved, and the proper Young diagram product rules explained 
earlier.  In any case, we retain only the lowest-order contributions to each 
matrix element.

We now turn to the $M_{\rm D}$ mass matrix which connects the ({\bf 10}) 
lefthanded down quarks and the $({\bf \overline{5}})$ lefthanded conjugate 
quarks with either the $({\bf \overline{5}})\higgs{495}$ or 
$({\bf \overline{5}})\higgsbar{12}$ EW Higgs.  The same considerations 
will apply to the $M_{\rm L}$ mass matrix connecting the $({\bf \overline{5}})$
lefthanded charged leptons and the $({\bf 10})$ lefthanded charged conjugate 
leptons, where the diagrams are the transpose of those for the down quarks.
Whether or not dim-4 contributions appear in these mass matrices depends on 
which of the two possible down-type Higgs are chosen for the models to be 
illustrated, $({\bf \overline{5}})\higgs{495}$ or 
$({\bf \overline{5}})\higgsbar{12}$, as well as the light family 
assignments.  In the first instant, a dim-4 vertex may be present involving 
\begin{equation}\label{eq:dim4Dij}
{\bf Dij}:\quad ({\bf 10}){\bf \overline{220}_i} . 
  ({\bf \overline{5}})\higgs{495} . ({\bf \overline{5}}){\bf \overline{12}_j},\\
\end{equation}
while in the second instance, a dim-4 vertex may be present involving either 
of the two possibilities
\begin{equation}\label{eq:dim4D3j}
\begin{array}{rl}
{\bf D3j}:\qquad & ({\bf 10}){\bf 66_3} . ({\bf \overline{5}})\higgsbar{12} . ({\bf \overline{5}}){\bf \overline{12}_j},\\
{\bf Dij}:\qquad & ({\bf 10}){\bf \overline{220}_i} . 
  ({\bf \overline{5}})\higgsbar{12} . ({\bf \overline{5}}){\bf 495_j}.\\
\end{array}
\end{equation}
The transverse conditions apply for the corresponding {\bf Lji} Yukawa matrix 
elements.

Higher dimensional contributions can involve not only ({\bf 10}) and 
$({\bf \overline{10}})$ intermediate states as for the $M_{\rm U}$ mass matrix, 
but also ({\bf 5}) and $({\bf \overline{5}})$ states.  Again from 
Eq.~\eqref{eq:massivefermions}, we see the latter choices are just 
$({\bf 5}){\bf 12} \times ({\bf \overline{5}}){\bf \overline{12}}$ 
and $({\bf 5}){\bf \overline{495}} \times ({\bf \overline{5}}){\bf 495}$.  The 
same considerations as in the previous paragraphs also apply for the ({\bf 1}) 
and ({\bf 24}) Higgs vertices.

For the $M_{\rm DN}$ mass matrix connecting the \SU{5} $({\bf \overline{5}})$ 
lefthanded neutrinos with the ({\bf 1}) lefthanded conjugate neutrinos, we 
assume that the same ({\bf 5})\higgsbar{495} EW Higgs is involved as 
for the up quark sector.  A dim-4 contribution to the $M_{\rm DN}$ mass matrix 
is possible, if one of the (${\bf \overline{5}}$) family states arises from the 
${\bf 495}$ as in the second family assignment of Eq.~\eqref{eq:familyids}, 
but not so otherwise.
For the higher dimensional contributions to the $M_{\rm DN}$ mass 
matrix, the massive fermion insertions involve the same ({\bf 5}) and 
$({\bf \overline{5}})$ possibilities as for the $M_{\rm D}$ and $M_{\rm L}$ mass 
matrices.  As with the other three Dirac matrices, singlet ({\bf 1}) and 
adjoint ({\bf 24}) Higgs scalars can appear in the $M_{\rm DN }$ matrix elements.
 
Finally, for the $M_{\rm MN}$ heavy righthanded Majorana mass matrix, since 
only $({\bf 1}){\bf 1}$ fermion singlets are involved, any mass insertions must 
involve only ({\bf 1}) singlets.  This
fact then negates the appearance of $({\bf 24})\higgs{5148}$ or 
$({\bf 24})\higgsbar{5148}$ Higgs contributions which would allow 
complex VEVs.  Hence the Majorana matrix in all models discussed here will be 
real.  The simplest dim-4 mass contribution is then given by 
\begin{equation}\label{eq:dim4MNij}
{\bf MNij}:\quad ({\bf 1}){\bf 1_i} . ({\bf 1}){\bf 1_H} . 
                ({\bf 1}){\bf 1_j}\\
\end{equation}
for all $i$ and $j$.  Note that this Higgs singlet must carry lepton number 
$L = 2$, in 
order to balance the two $L = - 1$ lefthanded conjugate neutrino assignments.
When this Higgs singlet obtains a VEV, $L$ is broken by two units, and the 
Majorana mass matrix element obtains a mass $\Lambda_R$.  The mass matrix 
then corresponds to a democratic matrix, aside from $O(1)$ prefactors 
which make the matrix non-singular.

In general, a restricted set of Higgs singlets and/or massive fermions may 
provide just one contribution to each mass matrix element.  Allowing more 
and more Higgs singlets and massive fermion insertions may lead to many
contributions of the same, higher, or even lower order for certain matrix 
elements.  Since only the lowest-dimensional contributions per matrix element 
are of interest, the more contributing tree diagrams that appear, the flatter 
the hierarchy will tend to be for any given mass matrix.  

\subsection{Illustrated Structure for One Model of Interest}

We have selected one model leading to interesting mixing results as a way of 
illustrating the steps involved to form the mass matrices and their consequent
mixing matrices and mixing parameters.  The model in question has the following
family structure, massive fermions, and Higgs fields:\\
\begin{equation}\label{eq:FermionAssignments}
\begin{array}{llll}
  \text{First Family:}\quad & (\irrep{10}){\bf 495_1} \rightarrow u_\text{L}, 
   u^c_\text{L}, d_\text{L}, e^c_\text{L};\quad & (\irrepbar{5}){\bf \overline{220}_1} \rightarrow d^c_\text{L}, e_\text{L}, \nu_{e\text{L}};\quad
     & (\irrep{1}){\bf 1_1} \rightarrow N^c_{1\text{L}}\\[1mm]
  \text{Second Family:}\quad & (\irrep{10}){\bf \overline{220}_2} 
     \rightarrow c_\text{L}, c^c_\text{L}, s_\text{L}, \mu^c_\text{L};\quad 
     & (\irrepbar{5}){\bf \overline{12}_2}
     \rightarrow s^c_\text{L}, \mu_\text{L}, \nu_{\mu\text{L}};\quad 
     & (\irrep{1}){\bf 1_2} \rightarrow N^c_{2\text{L}}\\[1mm]
  \text{Third Family:}\quad & (\irrep{10}){\bf 66_3} \rightarrow t_\text{L}, 
     t^c_\text{L}, b_\text{L}, \tau^c_\text{L};\quad 
     & (\irrepbar{5}){\bf \overline{12}_3} 
     \rightarrow b^c_\text{L}, \tau_\text{L}, \nu_{\tau\text{L}};\quad
     & (\irrep{1}){\bf 1_3} \rightarrow N^c_{3,\text{L}}
    \end{array}\quad\;
\end{equation}

\begin{equation}
 \begin{array}{ll}
     \text{Massive fermions:}\qquad & \irrep{12}{\times} \irrepbar{12},\quad
       \irrep{66}{\times} \irrepbar{66},\quad
       \irrep{220}{\times} \irrepbar{220},\quad 
       \irrep{495}{\times} \irrepbar{495},\quad 
       \irrep{792}{\times} \irrepbar{792} \\ 
     \text{Higgs bosons:}\qquad & (\irrep{5})\irrepbarsub{495}{H},  
       (\irrepbar{5})\irrepbarsub{12}{H}, (\irrep{24})\irrepsub{5148}{H}, 
       (\irrep{24})\irrepbarsub{5148}{H}, (\irrep{1})\irrepsub{1}{H},\\ 
     & (\irrep{1})\higgs{66}, (\irrep{1})\higgsbar{66},
       (\irrep{1})\higgs{792}, (\irrep{1})\higgsbar{792} \\[0.1in]
 \end{array}
\end{equation}

From the above irreps appearing in the model, we can construct the 
leading-order contributions to each Yukawa matrix element.  The complete list 
for this model is presented in Appendix B. for the ${\bf U},\ {\bf D},\ 
{\bf L},\ {\bf DN}$, and ${\bf MN}$ matrix elements.  For the ${\bf U}$ 
Yukawa matrix, dimensional contributions of order 4, 5, and 6 are found to 
appear, which are scaled according to the ratios 1 : $\epsilon$ : $\epsilon^2$, 
where $\epsilon$ is related to the ratio of the \SU{5} scale to the \SU{12} 
scale.  More precisely, $\epsilon$ is set equal to the ratio of a singlet
VEV, times its fermion coupling, divided by the \SU{12} unification scale 
where the massive fermions obtain their masses.  The ${\bf 5}$ and 
${\bf \bar{5}}$ EW VEVs are labeled $v_u$ and $v_d$, 
respectively, while the $({\bf 1}){\bf 1_H}$ $\Delta L = 2$ VEV is set equal 
to $\Lambda_R$.  The VEVs for $({\bf 24})\higgs{5148}$ and 
$({\bf 24})\higgsbar{5148}$ involve $\kappa$ and $\kappa^*$, respectively, 
as noted earlier in Eq.~(11).

The five mass matrices for the model in question then are found to have the 
following textures:  
\begin{displaymath}
    \begin{aligned}
       M_\text{U} =&
       \begin{pmatrix}
           \hu{11}\epsilon^2 & \hu{12}(\epsilon^2 - \frac{2\kappa\epsilon}{3})
               & \hu{13}\epsilon \\
           \hu{21}(\epsilon^2 + \frac{2\kappa\epsilon}{3}) & 
             \hu{22}(\epsilon^2 - \frac{4\kappa^2}{9}) &
             \hu{23}(\epsilon + \frac{2\kappa}{3}) \\
           \hu{31}\epsilon & \hu{32}(\epsilon - \frac{2\kappa}{3}) & \hu{33} \\
       \end{pmatrix} v_u, \\
       M_\text{D} =&
       \begin{pmatrix}
           2\hd{11}\epsilon^2 & \hd{12}\epsilon & \hd{13}\epsilon \\
           \hd{21}\epsilon & 2\hd{22}\epsilon & 2\hd{23}\epsilon \\
           \hd{31}\epsilon & \hd{32} & \hd{33} \\
       \end{pmatrix} v_d, \\
       M_\text{L} =&
       \begin{pmatrix}
           2\hl{11}\epsilon^2 & \hl{12}\epsilon & \hl{13}\epsilon \\
           \hl{21}\epsilon & 2\hl{22}\epsilon & \hl{23} \\
           \hl{31}\epsilon & 2\hl{32}\epsilon & \hl{33} \\
       \end{pmatrix} v_d, \\
    \end{aligned}
\end{displaymath}
\begin{equation}
    \begin{aligned}
       M_\text{DN} =&
       \begin{pmatrix}
           2\hdn{11}\epsilon & 2\hdn{12}\epsilon & 2\hdn{13}\epsilon \\
           \hdn{21}(2\epsilon - \kappa) & \hdn{22}(2\epsilon - \kappa) 
               & \hdn{23}(2\epsilon - \kappa) \\
           \hdn{31}(2\epsilon - \kappa) & \hdn{32}(2\epsilon - \kappa) 
               & \hdn{33}(2\epsilon - \kappa) \\
       \end{pmatrix} v_u, \\
       M_\text{MN} =&
       \begin{pmatrix}
           \hmn{11} & \hmn{12} & \hmn{13} \\
           \hmn{21} & \hmn{22} & \hmn{23} \\
           \hmn{31} & \hmn{32} & \hmn{33} \\
       \end{pmatrix} \Lambda_R.\\[0.1in]
    \end{aligned}
\end{equation}

\noindent The corresponding $h$'s are the prefactors to be determined 
numerically and are all required to lie in the range $\pm [0.1, 10]$ to 
achieve a satisfactory model that avoids fine tuning.  Note that 
$M_\text{U}$ exhibits a hierarchical structure, $M_\text{DN}$ and $M_\text{MN}$ 
do not, while $M_\text{D}$ and $M_\text{L}$ have no simple hierarchical 
structure.

\section{MODEL SCAN AND FITTING PROCEDURE}

In this section we explain the aforementioned computerized model scan in more 
detail. The scan determines anomaly-free sets of family assignments for 
$\SU{N}$ irreps and scans possible unification models by adding EW Higgs fields
and $\SU{5}$ Higgs singlets, as well as sets of massive fermions in a 
systematic way. The scan is built on top of LieART for the determination of 
tensor products extended to handle products of embeddings as described in 
Sect.~\ref{sec:SU12UnificationModels}. Potential models are fit to 
phenomenological particle data, such as masses, mixing angles and phases, to 
analyze their viability. The scan is not restricted to \SU{12} or a specific 
anomaly-free set of family assignments as discussed in this article, but we 
found \SU{12} to be the lowest rank yielding realistic models not requiring 
discrete group extensions of the symmetry, and its lowest anomaly-free set of 
irreps is maximally economical as it assigns all \SU{12} irreps to \SU{5} 
family irreps.

A pure brute-force scan of all possible family assignments and sets of Higgs 
and massive fermions has proven impractical due to the enormous number of 
possible combinations. Instead, we break up the full number of combinations 
into independent parts that are organized in enclosed loops: (1) Fermions  
embedded in \irrep{10}'s of \SU{5}, which include all up-type quarks, are first
assigned to suitable chiral irreps of the \SU{12} anomaly-free set and prove 
sufficient to construct the 
$M_{\rm U}$ mass matrix, once the sets of Higgs and massive fermion irreps 
are defined. (2) Likewise, assignment of fermions embedded in 
${\bf \overline{5}}$ of \SU{5} complete the quark and charged lepton sectors 
and allows one to compute the $M_{\rm D}$ and $M_{\rm L}$ mass matrices and thus 
the CKM matrix. We fit the $M_{\rm U}$, $M_{\rm D}$ and $M_{\rm L}$ mass matrix 
prefactors and four of the model parameters to the known quark masses and 
mixing angles, as well as charged lepton masses, at the GUT scale according to 
\cite{Bora:2012tx}. (3) Only for 
viable quark models do we loop over assignments of \SU{12} irreps embedding 
\SU{5} singlets as Majorana neutrinos. These assignments allow the construction
 of the $M_{\rm DN}$ and $M_{\rm MN}$ mass matrices and thus a fit to the lepton 
sector phenomenology. To this end we fit the $M_{\rm DN}$ and $M_{\rm MN}$ 
prefactors, as well as the righthanded scale $\Lambda_R$, to the known 
neutrino mass squared differences and two PMNS mixing angles. The 
$M_{\rm U}$, $M_{\rm D}$ and $M_{\rm L}$ prefactors and all other parameters 
from the quark sector remain fixed as determined by the first fit to avoid the 
variation of too many fit parameters at once. The lepton sector fit is 
performed twice: one favoring normal and the other inverted hierarchy of the 
light neutrino masses. Further details follow below.

\subsection{Scan of Assignments}

First, a list of anomaly-free sets of totally antisymmetric \SU{N} irreps that 
yield three families on the \SU{5} level is constructed, where $N > 5$.  The 
list is ordered by 
the total number of \SU{N} irreps in the sets and, since there is an infinite 
number of anomaly-free sets, is cut off at some chosen maximum. For \SU{12} a 
list of the simpler anomaly-free sets has been given in \eqref{eq:afsets}. In 
looping over this list, the scan performs family assignments only for irreps 
from one set at a time to ensure freedom from anomalies.

For each anomaly-free set the scan loops over the \SU{12} irreps containing 
\irrep{10}'s of \SU{5} for the assignment of the three up-type quarks to 
construct the $M_{\rm U}$ mass matrix. In terms of Young tableaux the 
\irrep{10}'s are embedded in the upper part of the column for the \SU{12} 
irreps, i.e., the regular embedding. Similarly, the scan loops over the \SU{12}
irreps containing \irrep{\overline{5}}'s of \SU{5} for the assignment of the 
three down-type quarks and leptons in a later step.

In a third loop the scan constructs subsets of possible assignments of EW Higgs
doublets, \SU{5} Higgs singlets, and massive fermion pairs. Both, the \SU{12} 
Higgs irreps and the massive fermion pairs are selected from all totally 
antisymmetric complex irreps with the \SU{5} EW Higgs and Higgs singlet irreps 
being regularly embedded. For our special \SU{12} scenario at hand we add the 
$(\irrep{24})\higgs{5148}$ and $(\irrep{24})\higgsbar{5148}$ to accommodate a 
CP phase and to abet the breaking of \SU{5} to the SM. To reduce the number 
of Higgs sets from the beginning, we keep only those EW Higgses that yield a 
dim-4 mass term for the ${\bf U33}$ element with the selected third-family 
fermion assignment, i.e., the largest contribution to the top-quark mass term 
at lowest order, as pointed out in 
Sect.~\ref{ssec:MassMatrixElementConstruction}.  For the simple anomaly-free 
set of \SU{12} the only possible ${\bf U33}$ at dim 4 using 
the regular embedding is given in expression \eqref{eq:U33}. The loop over 
Higgs and massive-fermion-pair subsets starts with the smallest set of Higgses 
and massive fermions increasing to larger ones. Limits on the subset size can 
be imposed to focus on economical models.

With the assignments of the \SU{12} irreps containing the \irrep{10}'s of 
\SU{5}, the \irrepsub{5}{H}'s associated with EW doublets, and \SU{5} Higgs 
singlets, as well as massive fermions assigned to \SU{12} irreps,  
the $\mathbf{U}$ matrix elements can be constructed. For a given set of 
fermion, Higgs and massive-fermion assignments determined by the iteration of 
the enclosing loops, the scan tries to construct diagrams for each 
matrix element beginning with a minimum, dim-4 or higher.  If none is found at
some dimension, it tries a higher dimension up to an adjustable upper limit. 
If one or more 
diagrams for a given dimension is found, the scan will turn to the next 
matrix element. Thus, only the lowest order contribution is taken into account.
The algorithm allows one to set a range of admissible dimensions for each 
matrix element, e.g., the ${\bf U11}$ element must not be of dimension 4 or 5, 
but may be of dimension 6 or 7. It is also possible to allow for no 
contribution 
up to a maximum dimension, i.e., there may be no contribution at all amounting 
to a texture zero or a contribution of an even higher dimension, which is 
not analyzed further. A mass-term diagram is constructed from Higgs and massive 
fermion insertions depending on its dimension. The validity of the constructed 
mass-term diagrams is ensured if all vertices are singlets on their own at 
both the \SU{12} and \SU{5} levels and under application of the Young-tableaux 
multiplication rules. A mass-term diagram can then be translated to powers of 
$\epsilon$ and $\kappa$ according to the orders of singlet VEVs and VEVs of 
$(\irrep{24})\higgs{5148}$ and $(\irrep{24})\higgsbar{5148}$, respectively.

With $M_{\rm U}$ mass matrices matching the desired texture set by the 
dimension requirements, the $M_{\rm D}$ mass matrix is constructed from 
subsets of three unassigned irreps of the anomaly-free set containing 
\irrepbar{5}'s of \SU{5}, looping over the regular embedding. The 
construction of the mass matrix elements is analogous to that for the 
$M_{\rm U}$ mass matrix. The $M_{\rm L}$ matrix can be constructed from the 
reverse of the ${\bf D}$ matrix element diagrams. With the $M_{\rm D}$ and 
$M_{\rm L}$ matrices constructed, all assignments of the quark and charged 
lepton sectors are fixed. 

We fit the quark sector and the charged leptons to phenomenological data run 
to the GUT scale taken from \cite{Bora:2012tx}. The description of this fit
and the lepton sector fit is deferred to the next section.  Since the quark 
sector is fully determined without the assignment of lefthanded conjugate 
Majorana neutrinos, we detach quark and lepton sector fits, to avoid fitting 
seemingly complete models where the quark sector itself does not reproduce SM 
phenomenology.

For models with quark and charged lepton sectors determined to be viable by 
the fit, a last loop over subsets of irreps assigned to Majorana neutrinos is 
performed. This requires \SU{12} irreps containing \SU{5} singlets. They are 
taken from unassigned irreps of the anomaly-free set or from additional \SU{12}
singlets, since they do not need to be chiral. The $M_{\rm DN}$ and 
$M_{\rm MN}$ matrices are constructed in analogy with the $M_{\rm U}$ and 
$M_{\rm D}$ matrices. Once they are known, the lepton sector can be fit as 
well using the fit results of the quark sector performed in the stage prior to 
the assignment of Majorana neutrinos.

\subsection{Quark and Lepton Sector Fits}

Now we return from a more general description of the scanning 
procedure to our specific model setup to describe the separate quark and 
lepton sector fits to phenomenological data using the simplest anomaly-free set 
of \SU{12} and the addition of $(\irrep{24})\higgs{5148}$ and 
$(\irrep{24})\higgsbar{5148}$ scalars with complex valued VEVs introducing a 
source of CP violation.

\subsubsection{Quark Sector Fit}

The $M_{\rm U},\ M_{\rm D},$ and $M_{\rm L}$ matrices enter the quark and charged
lepton fit in terms of their prefactors $\hu{ij}$, $\hd{ij}$ and $\hl{ij}$, 
powers of $\epsilon$ related to the \SU{5} singlet VEVs appearing, and the 
complex-valued VEVs of $(\irrep{24})\higgs{5148}$ and 
$(\irrep{24})\higgsbar{5148}$ involving $\kappa$ and $\kappa^*$, respectively. 
The two EW VEVs of the 2-Higgs-Doublet-Model are 
labeled $v_u$ and $v_d$, with only one independent and chosen to be $v_u$ 
since $v^2=v_u^2+v_d^2$ must give $v=174$\,GeV.  Because  
$(\irrep{24})\higgs{5148}$ and $(\irrep{24})\higgsbar{5148}$ give different 
contributions to the $M_{\rm D}$ and $M_{\rm L}$ mass matrices according to 
\eqref{eqn:5148VEVs} and asymmetric contributions to the $M_{\rm U}$ matrix, we 
refrain from imposing any symmetries on the prefactors and allow them to remain
independent parameters. Thus, we have 
27 real prefactors (9 per mass matrix), one real ratio $\epsilon$, one 
complex ratio $\kappa$, and the EW VEV $v_u$, yielding a total of 31 
parameters for the quark sector fit. 

As initial values of the fit parameters we choose 
$\epsilon{=}|\kappa|=1/6.5^2{=}0.0237$, motivated by \cite{Babu:1999me}, 
$\arg(\kappa)=45^\circ$ and $v_u = \sqrt{v^2/(1 + \epsilon^2)}$, where 
$v=174$\,GeV. The choice of unity for all initial values of the 
prefactors leads to cancellations in the matrix diagonalizations, thus 
resulting in fine tuning. Hence, we choose to set initial prefactor values 
randomly in the intervals $[-1.3, -0.7]$ and $[0.7, 1.3]$.  Models with any 
prefactor fit to absolute values lower than |0.1| or higher than |10| 
are discarded. Fits with other randomly assigned prefactor initial values for
such a model are tried until we either find a successful fit, or after a 
certain number of trials have been performed without success, we discard the 
model.

We perform the fit against phenomenological data at the \SU{5} unification 
scale using values for the six quark and three charged lepton masses from 
\cite{Bora:2012tx}. We use the measured values of the three quark mixing 
angles and phase. The renormalization group flow of the CKM matrix is governed 
by the Yukawa couplings, which are small except for the top quark. According to 
\cite{Balzereit:1998id} the running of the matrix elements of the first two 
families is negligible and small for the third family. Thus we have neglected 
the running of the quark mixing angles and phase. In total we use 13 
phenomenological data points.

The phenomenological implications of the models are compared with data by  
diagonalizing the mass matrices to obtain the quark and charged lepton 
masses and determine the CKM matrix from the unitary transformations 
diagonalizing $M_{\rm U}$ and $M_{\rm D}$. By transforming the CKM matrix into 
the standard parametrization, the three mixing angles and the CKM phase are 
easily obtained, as we explain in the following.

Since the Dirac matrices $M_{\rm U},\ M_{\rm D}$ and $M_{\rm L}$ are generally
not Hermitian, we form their lefthanded Hermitian products and diagonalize 
them with lefthanded rotations to obtain positive real eigenvalues as squares 
of the corresponding masses: 
\begin{equation}\label{eqn:diagonalization}
    \begin{aligned}
        U_{\rm U}^\dagger M_{\rm U} M_{\rm U}^\dagger U_{\rm U} &=
                   \diag(m_u^2, m_c^2, m_t^2), \\
        U_{\rm D}^\dagger M_{\rm D} M_{\rm D}^\dagger U_{\rm D} &=
                   \diag(m_d^2, m_s^2, m_b^2), \\
        U_{\rm L}^\dagger M_{\rm L} M_{\rm L}^\dagger U_{\rm L} &=
                   \diag(m_e^2, m_\mu^2, m_\tau^2).\\
    \end{aligned}
\end{equation}
The Cabibbo-Kobayashi-Maskawa (CKM) matrix $V_\text{CKM}$ encodes the mismatch 
of the mass and flavor eigenstates of the up- and down-type quarks and is 
calculated from the unitary transformations $U_{\rm U}$ and $U_{\rm D}$: 
\begin{equation}\label{eqn:CKMMatrixFromUnitaryTrasformations}
    V_\text{CKM} = U_{\rm U}^\dagger U_{\rm D}.
\end{equation}
The CKM matrix in standard parametrization of the Particle Data Group 
\cite{PDG:2014} with  $c_{ij}=\cos\theta_{ij}$ and $s_{ij}=\sin\theta_{ij}$ 
is given by 
\begin{equation}
      V_\text{CKM}=  \begin{pmatrix}
         c_{12}c_{13}       &     s_{12}c_{13}  &  s_{13}e^{-i\delta}  \\
         -s_{12}c_{23}-c_{12}s_{23}s_{13}e^{i\delta} &  
              c_{12}c_{23}-s_{12}s_{23}s_{13}e^{i\delta}  &  s_{23}c_{13}        \\
         s_{12}s_{23}-c_{12}c_{23}s_{13}e^{i\delta}  & 
              -c_{12}s_{23}-s_{12}c_{23}s_{13}e^{i\delta}  &  c_{23}c_{13}        \\
                       \end{pmatrix}.
\end{equation}

A CKM matrix obtained by \eqref{eqn:CKMMatrixFromUnitaryTrasformations} can be 
brought into standard form by redefining five relative phases of quark fields, 
that are unphysical, or by extracting the three angles and the phase directly. 
The angles can be obtained from
\begin{equation}
      \theta_{12} = \arctan\left(\frac{|V_{12}|}{|V_{11}|}\right), \quad 
      \theta_{23} = \arctan\left(\frac{|V_{23}|}{|V_{33}|}\right) \quad 
      \text{and} \quad
      \theta_{13} = \arcsin\left(|V_{13}|\right).\\
\end{equation}
To determine the phase we perform a phase rotation of columns one and two such 
that $V^\prime_{11}$ and $V^\prime_{12}$ become real, where the prime denotes 
the rotated columns. We equate the quotient of the $V^\prime_{22}$ and 
$V^\prime_{21}$ elements with the corresponding expression in the standard form 
\renewcommand{\exp}[1]{\ensuremath{\text{e}^{#1}}}
\begin{equation}
      r=\frac{V^\prime_{22}}{V^\prime_{21}} = \frac{V_{22}}{V_{21}}
            \exp{i(\phi_{11}-\phi_{12})} = \frac{c_{12}c_{23}-s_{12}s_{23}s_{13}
            \exp{i\delta}}{-s_{12}c_{23}-c_{12}s_{23}s_{13}\exp{i\delta}} ,
\end{equation}
where $\phi_{11}$ and $\phi_{12}$ are the phases of $V_{11}$ and $V_{12}$, 
respectively, and solve for the phase $\delta$ yielding
\begin{equation}
      \delta = \arg\left(\frac{c_{12}c_{23}+rs_{12}c_{23}}
             {s_{12}s_{23}s_{13}-rc_{12}s_{23}s_{13}}\right).
\end{equation}

\subsubsection{Lepton Sector Fit}

Only quark models with a reasonably good fit are extended to include 
assignments of the lefthanded conjugate Majorana neutrinos in a loop over 
all their possibilities. 
For the simplest anomaly-free \SU{12} model of interest here, since all six 
non-trivial irreps have been assigned to the \SU{5} ${\bf 10}$ and ${\bf 
\overline{5}}$ family irreps, the three heavy neutrinos are all assigned to 
\SU{12} singlets. The $M_{\rm DN}$ and $M_{\rm MN}$ matrices are then determined,
and the complex symmetric light-neutrino mass matrix is 
obtained via the Type I seesaw mechanism,
 \begin{equation}
    M_\nu = -M_\text{DN}M_\text{MN}^{-1}M_\text{DN}^T.
\end{equation}

By convention, the complex symmetric $M_\nu$ matrix is to be diagonalized by
the unitary transformation 
\begin{equation}
\label{eq:mnueveq}
    U^T_\nu M_\nu U_\nu = {\rm diag}(m_1, m_2, m_3),
\end{equation}
to yield positive real eigenvalues $m_i$.  This requires a very special 
unitary $U_\nu$ transformation, for in general the eigenvalues will be complex.
To acquire the desired result, we form the Hermitian product 
$M^\dagger_\nu M_\nu$ and perform the unitary transformation by using (33),
\begin{equation}
\label{eq:mnu2eveq} 
        U_\nu^\dagger M_\nu^\dagger M_\nu U_\nu = \diag(m_1^2, m_2^2, m_3^2),
\end{equation}
to obtain positive real eigenvalues, $m_i^2$, and the transformation matrix 
$U_\nu$.  Clearly, Eq.~\eqref{eq:mnu2eveq} is invariant to a phase 
transformation
$\Phi$ from the right together with its conjugate phase transformation from
the left.  We now define $U'_\nu = U_\nu \Phi'$ to be the special unitary 
transformation, operating on $M_\nu$ as in Eq.~\eqref{eq:mnueveq}, which makes 
the neutrino mass eigenvalues real for the appropriate diagonal phase matrix 
$\Phi'$. 
The Pontecorvo-Maki-Nakagawa-Sakata (PMNS) matrix \cite{PMNS}, $V_\text{PMNS}$,
follows from $U'_\nu$ and the unitary transformation $U_\text{L}$ diagonalizing 
the charged lepton mass matrix $M_\text{L}$, according to 
\begin{equation}
\label{eq:VPMNSorig}
    V_\text{\rm PMNS} = U_\text{L}^\dagger U'_\nu.
\end{equation}
The PDG phase convention \cite{PDG:2014} for the neutrino mixing matrix 
$U_{\rm PMNS}$ follows by phase transforming the left- and right-hand sides of 
$V_\text{PMNS}$ and then writing 
\begin{equation}
\label{eq:VPMNS}
    V_\text{PMNS} \equiv U_\text{PMNS} \Phi_\text{Majorana},
\end{equation}
where $\Phi_\text{Majorana} = {\rm diag}(e^{i\phi_1/2},\ e^{i\phi_2/2},\ 1)$ with 
Majorana phases $\phi_1$ and $\phi_2$ is the 
adjoint of the required righthanded phase transition matrix, so effectively
$U'_\nu$ is left untransformed from the right.  The neutrino mixing 
angles and Dirac phase are determined in analogy with the CKM matrix.

To accomplish the phase transformations of Eq.~\eqref{eq:VPMNSorig} in detail, 
we follow the procedure as for $V_{\rm CKM}$ in Eqs. (29) -- (31) to obtain 
$U_{\rm PMNS}$ in the PDG convention.    To restore the correct untransformed 
$U'_\nu$ as appears in (35), we then multiply by $\Phi^\dagger$ on the 
right to obtain Eq.~\eqref{eq:VPMNS} with $\Phi_{\rm Majorana} = \Phi^\dagger$. 

The effective mass $|\langle m_{ee}\rangle|$ for neutrinoless double beta 
decay \cite{PDG:2014} follows from Eq.~\eqref{eq:VPMNS} according to 
\begin{equation}
    |\langle m_{ee}\rangle| = |\Sigma_i\ (V_{\text{PMNS},{ei}})^2 m_i|,
    \qquad i = 1,2,3.
\end{equation}

\noindent This assumes that the light neutrino masses are the major contributors
to the corresponding loop diagrams for the effective mass contribution to 
neutrinoless double beta decay \cite{Rodejohann:2012}.

We fit the lepton sector with recent neutrino data \cite{Gonzalez-Garcia:2014}
for the mass squared 
differences of the light neutrinos $\abs{\Delta_{21}}$, $\abs{\Delta_{31}}$ and 
$\abs{\Delta_{32}}$ and the sines squared of the neutrino mixing angles,
$\sin^2\theta_{12}$ and $\sin^2\theta_{13}$. We do not fit to the Dirac CP 
phase or $\sin^2\theta_{23}$, but discard models that are not within the bounds 
of $0.34 \leq \sin^2\theta_{23} \leq 0.66$, since values from current global 
fits of neutrino data can only provide this range or smaller. In total we fit 
to five data points.
Since the $M_{\rm MN}$ matrix is symmetric, and involves the righthanded scale 
$\Lambda_R$ as an additional parameter and the $M_{\rm DN}$ matrix is not 
symmetric, 
the lepton sector fit encompasses 16 fit parameters. Initial values for the 
prefactors are chosen in analogy to the quark sector fit, and we set the 
initial value of the righthanded scale to $\Lambda_R=4\times10^{14}$\,GeV. 
Models with prefactors not within the range $\pm [0.1, 10]$ are discarded and 
refit with other randomly assigned prefactors, as was done for the fits of 
quark sector models. The leptonic fits for satisfactory models are carried out
favoring first normal hierarchy (NH) and then inverted hierarchy (IH). In some 
cases, satisfactory models for both hierarchies can be obtained with the same 
set of mass matrix textures, but with different sets of prefactors of course.\\

\begin{table}[h]
\begin{center}
\begin{tabular}{crclc}
\hline\hline
  &\multicolumn{3}{c}{\bf Quark and Lepton Data} &\qquad {\bf Fitted Results} \\
\hline\hline
${m_u\ ({\rm MeV})}$ & 0.3963 &$\pm$& 0.1395 & 0.3950 \\
${m_c\ ({\rm GeV})}$ & 0.1932 &$\pm$& 0.0243   & 0.1932  \\
${m_t\ ({\rm GeV})}$ & 80.4472 &$\pm$& 2.7643    & 80.45  \\
${m_d\ ({\rm MeV})}$ & 0.9284 &$\pm$& 0.3796  & 0.9143  \\
${m_s\ ({\rm MeV})}$ & 17.6097 &$\pm$& 4.7855  & 17.60  \\
${m_b\ ({\rm GeV})}$ & 1.2424 &$\pm$& 0.0599  & 1.243  \\
${m_e\ ({\rm MeV})}$ &\hspace{0.35in} 0.3569 &$\pm$& 0.0003 & 0.3509 \\
${m_\mu\ ({\rm MeV})}$ & 75.3570 &$\pm$& 0.0713 & 75.42 \\
${m_\tau\ ({\rm GeV})}$ & 1.6459 &$\pm$& 0.0160 & 1.646 \\
\hline
${\theta_{12}^q}$ & 13.04 &$\pm$&$0.05^\circ$ & 13.04$^\circ$ \\
${\theta_{23}^q}$ & 2.38 &$\pm$&$0.06^\circ$ & 2.381$^\circ$ \\
${\theta_{13}^q}$ & 0.201 &$\pm$&$0.011^\circ$ & 0.2037$^\circ$ \\
${\delta^q}$     & 68.75 &$\pm$&$4.584^\circ$ & 68.76$^\circ$ \\[0.1in]
\hline
${\Delta_{21}\ (10^{-5} eV^2)}$ & 7.50 &$\pm$& 0.18 & 7.4  \\
${|\Delta_{31}|\ (10^{-3} eV^2)}$ & 2.45 &$\pm$& 0.047 & 2.5 (2.4) \\
${|\Delta_{32}|\ (10^{-3} eV^2)}$ & 2.45 &$\pm$& 0.047 & 2.4 (2.5) \\
${\sin^2 \theta_{12}}$ & 0.304 &$\pm$& 0.012 & 0.304 \\
${\sin^2 \theta_{13}}$ & 0.0218 &$\pm$& 0.001 & 0.0218 \\
\hline\hline
\end{tabular}
\end{center}
\caption{\label{tab:DatatableI} Phenomenological data with masses
  at the GUT scale and fitted model results for the special case illustrated.  
  The NH (IH) results are indicated without (with) parentheses for 
  $|\Delta_{31}|$ and $|\Delta_{32}|$.  The quark data are taken from Ref. 
  \cite{Bora:2012tx} and the neutrino data from \cite{Gonzalez-Garcia:2014}.}
\end{table}

\subsection{Fitting Results for Special Case Illustrated}

We begin with the known data, evaluated at the \SU{5} GUT scale, which will 
be fitted with the five model parameters and prefactors for the five mass
matrices.  For the quark and charged lepton sectors, this consists of the
nine masses and three CKM mixing angles and one phase listed in Table I.  
For the lepton sector, we make use of the three neutrino mass squared 
differences and two of the three neutrino mixing angles which
are also given in Table~I.  The unknown neutrino quantities then involve the 
mass hierarchy (MH), the 
righthanded Majorana scale $\Lambda_R$ fit parameter, the light and heavy 
neutrino masses, the octant and values of $\sin^2 \theta_{23}$ and $\delta$,
along with the Majorana phases, and the effective neutrinoless double beta 
decay mass.\\  

Following the above scanning and fitting procedures for the special case 
illustrated in Sect.~IV C., the following matrices have been obtained in 
terms of the parameters $\epsilon,\ \kappa,\ \kappa^*,\ v_u,\ v_d,\ \Lambda_R$ 
with the prefactors indicated explicitly.  For the quark and charged lepton 
mass matrices the results obtained are 
\begin{equation}
    \begin{aligned}
       M_\text{U} =&
       \begin{pmatrix}
           -0.47\epsilon^2 & 0.48(\epsilon^2 - \frac{2\kappa\epsilon}{3})
               & 0.28\epsilon \\
           -1.0(\epsilon^2 + \frac{2\kappa\epsilon}{3}) & 
             1.9(\epsilon^2 - \frac{4\kappa^2}{9}) &
             0.61(\epsilon + \frac{2\kappa}{3}) \\
           0.67\epsilon & 2.5(\epsilon - \frac{2\kappa}{3}) & -0.46 \\
       \end{pmatrix} v_u, \\
        M_\text{D} =&
       \begin{pmatrix}
           -2.8\epsilon^2 & 1.1\epsilon & 0.15\epsilon \\
           0.84\epsilon & 5.3\epsilon & 0.2\epsilon \\
           -1.3\epsilon & -0.97 & -0.15 \\
       \end{pmatrix} v_d, \\
       M_\text{L} =&
       \begin{pmatrix}
           1.8\epsilon^2 & 0.85\epsilon & -1.2\epsilon \\
           0.55\epsilon & 4.6\epsilon & 0.37 \\
           -0.91\epsilon & -1.3\epsilon & -1.3 \\
       \end{pmatrix} v_d, \\
    \end{aligned}
\end{equation}

\noindent  with the parameters found to be  $\epsilon = 0.01453,\ 
\kappa = 0.02305\ e^{i27.53^\circ},\ v_u = 174\ {\rm GeV},\ v_d = 1.262\ 
{\rm GeV}$.

It turns out for this special model, both NH and IH solutions can be found.
With the parameters determined as above, the two sets of neutrino mass matrices 
are given by 
\begin{equation}
    \begin{aligned}
       M_\text{DN}^{\rm NH} =&
       \begin{pmatrix}
           -1.6\epsilon & 1.5\epsilon 
               & 3.3\epsilon \\
           -1.5(2\epsilon - \kappa) & 0.43(2\epsilon - \kappa) 
               & 0.76(2\epsilon - \kappa) \\
           -1.9(2\epsilon - \kappa) & 0.7(2\epsilon - \kappa) 
               & -1.9(2\epsilon - \kappa)\\
       \end{pmatrix} v_u, \\
       M_\text{MN}^{\rm NH} =&
       \begin{pmatrix}
           0.94 & -1.1 & -0.16 \\
           -1.1 & 1.5 & 1.3 \\
           -0.16 & 1.3 & -0.38 \\
       \end{pmatrix} \Lambda_R^{\rm NH}.
    \end{aligned}
\end{equation}

\noindent for NH and by 
\begin{equation}
    \begin{aligned}
       M_\text{DN}^{\rm IH} =&
       \begin{pmatrix}
           -2.8\epsilon & 2.4\epsilon & -1.9\epsilon \\
           1.3(2\epsilon - \kappa) & 1.3(2\epsilon - \kappa) 
               & 0.78(2\epsilon - \kappa) \\
           1.7(2\epsilon - \kappa) & 1.6(2\epsilon - \kappa) 
               & -1.6(2\epsilon - \kappa) \\
       \end{pmatrix} v_u, \\
       M_\text{MN}^{\rm IH} =&
       \begin{pmatrix}
           -1.2 & 0.26 & -0.4 \\
           0.26 & 0.62 & 2.0 \\
           -0.4 & 2.0 & 0.61 \\
       \end{pmatrix} \Lambda_R^{\rm IH}.
    \end{aligned}
\end{equation}
 
\noindent for IH,  where $\Lambda_R^{\rm NH} = 1.5 \times 10^{12}\ {\rm GeV}$ 
and $\Lambda_R^{\rm IH} = 1.3 \times 10^{12}\ {\rm GeV}$.

For the NH case the unknown neutrino mixing parameters and masses are 
determined to be 
\begin{equation}
\begin{array}{ccccc}
    \sin^2 \theta_{23} = 0.460, & \delta = - 121^\circ, &
    \phi_1 = -219^\circ, & \phi_2 = -67.4^\circ, & |\langle m_{ee}\rangle| = 
    2.45\ {\rm meV}\\
\multicolumn{5}{c}{
    m_1 = 2.76\ {\rm meV}, \qquad m_2 = 9.03\ {\rm meV}, 
          \qquad m_3 = 50.0\ {\rm meV},}\\
\multicolumn{5}{c}{
    M_1 = 5.86 \times 10^{11}\ {\rm GeV},\  M_2 = 1.76 \times 10^{12}\ 
          {\rm GeV}, \  M_3 = 4.35 \times 10^{12}\ {\rm GeV},}\\
\end{array}
\end{equation}
while for the IH case, 
\begin{equation}
\begin{array}{ccccc}
    \sin^2 \theta_{23} = 0.600, & \delta = - 50.2^\circ, & 
    \phi_1 = -319^\circ, & \phi_2 = 12.8^\circ, & |\langle m_{ee}\rangle| = 
    47.2\ {\rm meV},\\
\multicolumn{5}{c}{
    m_1 = 49.3\ {\rm meV},\qquad m_2 = 50.0\ {\rm meV},
          \qquad m_3 = 3.39\ {\rm meV},}\\
\multicolumn{5}{c}{    
    M_1 = 1.06 \times 10^{12}\ {\rm GeV},\ M_2 = 2.26 \times 10^{12}\ 
          {\rm GeV},\ M_3 = 3.33 \times 10^{12}\ {\rm GeV}.}\\
\end{array}
\end{equation}

\section{Results for Acceptable Models}

We now present the results for acceptable \SU{12} neutrino mixing models 
obtained with the scanning and fitting procedures outlined in Sect.~V. While 
in general we noted that two sets of EW Higgs doublets are available
for giving Dirac masses to the quarks and leptons, $({\bf 5})\higgs{12}$ and 
$({\bf 5})\irrepbarsub{495}{H}$ for the up-type quarks and Dirac neutrinos, and 
$({\bf \overline{5}})\irrepbarsub{12}{H}$ and $({\bf \overline{5}})\higgs{495}$
for the down-type quarks and charged leptons, only the Higgs doublet in the 
$({\bf 5})\irrepbarsub{495}{H}$ could provide a dim-4 contribution to the top 
quark mass.  Thus for simplicity we considered the $({\bf 5})\higgs{12}$ to 
contain an inert doublet and to be of no further interest.  On the other hand, 
both irreps leading to $({\bf \overline{5}})$ doublets seemed to be possible 
contributers to the EW VEVs of the down-type quarks and charged leptons.  But 
the full scan results to be displayed below have shown that only the 
$({\bf \overline{5}})\irrepbarsub{12}{H}$ appeared in successful models for the
simplest anomaly-free set.  Hence it suggests that we also consider the 
$({\bf \overline{5}})\higgs{495}$ to contain an inert Higgs doublet.

Concerning the permissible family assignments for the three $({\bf 10})$'s and 
the three $({\bf \overline{5}})$'s displayed in Eq. (18), along with their 
permutations, no satisfactory model appeared involving the second $({\bf 10})$ 
assignment, $({\bf 10})\irrepbarsub{220}{1}$, $({\bf 10})\irrepbarsub{220}{2}$,
$({\bf 10}){\bf 66_3}$.  The family assignments leading to acceptable models 
are labeled I and II for the two remaining ({\bf 10}) choices,
\begin{equation}
\begin{array}{rl}
  {\rm I}: & ({\bf 10}){\bf 495_1},\qquad ({\bf 10}){\bf \overline{220}_2},
    \qquad ({\bf 10}){\bf 66_3}\\
  {\rm II}: & ({\bf 10}){\bf \overline{220}_1},\qquad ({\bf 10}){\bf 495_2},
    \qquad ({\bf 10}){\bf 66_3}\\
\end{array}
\end{equation}
and A, B, and C for the three $({\bf \overline{5}})$ permutations,
\begin{equation}
\begin{array}{rlll}
  {\rm A}:&  ({\bf \overline{5}}){\bf \overline{12}_1}, & 
             ({\bf \overline{5}}){\bf \overline{12}_2}, &  
             ({\bf \overline{5}}){\bf \overline{220}_3} \\
  {\rm B}:&  ({\bf \overline{5}}){\bf \overline{12}_1}, &
             ({\bf \overline{5}}){\bf \overline{220}_2}, \qquad\quad &  
             ({\bf \overline{5}}){\bf \overline{12}_3} \\
  {\rm C}:&  ({\bf \overline{5}}){\bf \overline{220}_1}, \qquad\quad &
             ({\bf \overline{5}}){\bf \overline{12}_2}, & 
             ({\bf \overline{5}}){\bf \overline{12}_3} \\
\end{array}
\end{equation}

\input{ModelClass10C.tex}

Table II gives a summary of the models found acceptable by the scanning and 
fitting proceedure where the types of models are numbered according to their 
massive fermion content labeled MF1, MF2, etc. and their Higgs structure. For 
simplicity only the irreps are given with their conjugate irreps understood to 
be included.  The Higgs irreps include $({\bf 5})\irrepbarsub{495}{H}$, 
$({\bf \overline{5}})\irrepbarsub{12}{H}$, the $\Delta L = 2$ Higgs singlet
$({\bf 1})\irrepsub{1}{H}$, ({\bf 24})\irrepsub{5148}{H} and 
({\bf 24})\irrepbarsub{5148}{H} in all models.  In addition, it is found that 
proper models can be constructed either with the minimum number of Higgs 
singlets $({\bf 1})\irrepsub{792}{H}$ and $({\bf 1})\irrepbarsub{792}{H}$ in a 
few cases, or with these singlets plus the $({\bf 1})\irrepsub{66}{H}$ and 
$({\bf 1})\irrepbarsub{66}{H}$ pair 
in the majority of cases as indicated in the Table.  In fact, while seven 
other sets of Higgs singlets can be added to each model, the results are 
unmodified, for diagrams including those additional Higgs irreps all occur with 
higher-dimensional contributions to the matrix elements which we choose to 
neglect.  On the other hand, additional $\irrep{792}$ and $\irrepbar{792}$
fermions will add extra contributions to the (${\bf DNij}$) Yukawa matrix 
elements.

Also included in Table II are the fit parameters $\epsilon,\ |\kappa|$, and 
arg$(\kappa)$ which are adjusted to help give good fits to the charged lepton
masses and to the quark mass and mixing data.  The additional adjusted Higgs 
VEVs, $v_u$ and $v_d = \sqrt{174^2 - v_u^2}$, are found to lie in the range 
$v_d = 1 - 5$ GeV and $v_u \simeq 174$ GeV and are not included in the Table.
Recalling that the initial values for the fit parameters were chosen as 
$\epsilon = |\kappa| = 1/6.5^2 = 0.0237,$ and arg($\kappa$) = $45^\circ$, 
we see that the resulting fit parameters for $\epsilon$ and $|\kappa|$ are 
reasonably close to their starting values.   In particular, 
$|\kappa| \sim (0.5 - 5)|\epsilon|$ in most cases, so that the 
$({\bf 24})$ Higgs contributions, which serve to split the down quark 
and charged lepton spectra, and the Higgs singlet contributions are comparable.
To obtain a satisfactory model, we have
required that all prefactors lie in the range $\pm [0.1,10]$, i.e., within 
a factor of 10 of unity.  In most cases, the range is considerably tighter.
Not surprisingly, in all cases the quark mass and mixing parameters at the 
GUT scale can be fit accurately with the above model parameters and the 
overwhelming number of matrix element prefactors far exceeding the number of 
data points. 

\input{ModelNeutrino10C.tex}

To give a more complete picture of the results obtainable for successful 
models, we have made five separate complete runs of the scanning and fitting
procedure outlined above and labeled them by their run number in Table II.
Due to the Monte Carlo nature of the prefactor fitting, it is apparent from the 
table that no successful model assignments were obtained for all five runs, 
and in many cases for only two or three of the runs.  Nevertheless, the 
results are instructive.  

For succesful models found in run 4, we present in Table III the predictions 
for the neutrino mass and mixing 
parameters that were obtained with fits of $\Lambda_R$ and the known 
neutrino mass and mixing parameters, namely, the three $\Delta m^2_{ij}$'s
and the two sine squares of $\theta_{12}$ and $\theta_{13}$.  
In particular, we list for each model the neutrino mass hierarchy MH, 
$\Lambda_R$, the unknown heavy righthanded Majorana neutrino masses 
$M_1,\ M_2,\ M_3$, the light neutrino masses $m_1,\ m_2,\ m_3$, 
 $\sin^2 \theta_{23}$, the Dirac leptonic CP 
phase $\delta$, the Majorana phases $\phi_1$ and $\phi_2$, and the effective 
mass parameter $|\langle m_{ee}\rangle|$ for neutrinoless double beta decay.  
The latter prediction assumes the light neutrino masses are the major 
contributors to the corresponding loop diagrams.  Of course there are large 
spreads in the resulting predictions due to the Monte Carlo adjusted fit 
parameters and large number of matrix element prefactors.  In a number of 
family symmetry cases listed, both normal and inverted 
hierarchy models are acceptable with quite different sets of matrix element 
prefactors and $\Lambda_R$.  Of the 31 models found in run 4, it is 
interesting to note that 17 of them correspond to normal hierarchy, while 
14 have inverted hierarchy.  For all five runs with an average of 28 
successful models each, 76 have normal hierarchy while 64 have inverted 
hierarchy.  

In order to better grasp the distributions of results obtained in all five 
runs, we present several scatterplots.  In Fig. 1, $\delta$ is plotted vs. 
$\sin^2 \theta_{23}$ for 
normal hierarchy in (a) and for inverted hierarchy in (b).  The circles and 
squares for NH (upright and inverted triangles for IH) refer to the 
I~and~II~family ({\bf 10}) assignments, respectively, while the shadings 
distinguish the family $({\bf \overline{5}})$ assignments.   With regard to 
the $\sin^2 \theta_{23}$ distributions, the NH one slightly prefers the second
octant, while the IH one is equally split between the first and second octants.
It is apparent that most of the models favor small leptonic CP violation, for 
$\delta$ tends to lie near $0^\circ$ or $180^\circ$.  The IC6 run 5 model we 
have illustrated earlier with both NH and IH variations is among the 
exceptions.
\begin{figure}[floatfix, p!]
\begin{center}
  \includegraphics[width=6.5in,height=9in]{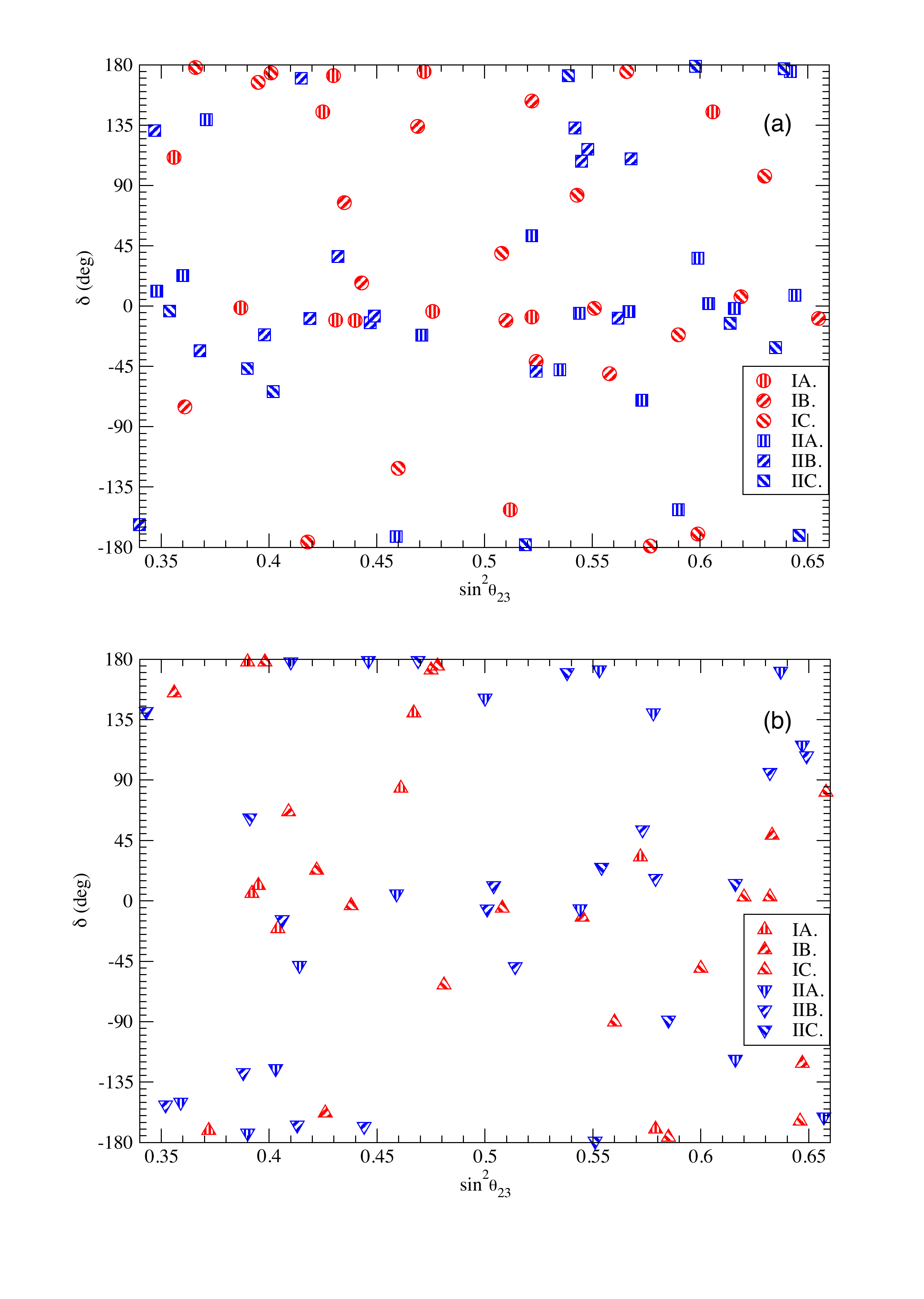}
  \vspace{-1in}
  \caption{Scatterplots of the CP $\delta$ vs. $\sin^2 \theta_{23}$ for 
    the NH models in (a) and for the IH models in (b).  The symbols label 
    the sets of models listed in Table II. for both normal and inverted 
    hierarchy.}
\label{fig:deltasinsq10.pdf}
\end{center}
\end{figure}

In Fig. 2 is displayed the effective mass parameter for neutrinoless
double beta decay vs. the lightest neutrino mass, $m_0 = m_1$ for NH and 
$m_0 = m_3$ for IH.  As expected the IH points lie higher than the NH ones.
The IH $|\langle m_{ee}\rangle|$ values cluster around 20 and 50 meV, while the 
NH ones generally fall below 10 meV with the smallest value occurring for 
$m_1 \sim 2$ meV.  The two clusterings occur because the difference of 
the two phases $\phi_1$ and $\phi_2$ for many of the IH models tends to be 
near $0^\circ$ or $180^\circ$.  From this figure we can conclude that the 
neutrinoless double beta decay experiments by themselves must be able to 
reach down to $m_{ee} = 10$ meV in order to rule out the inverse mass hierarchy 
in the framework of three righthanded neutrinos.  These results are well within 
the ranges found by the PDG in \cite{PDG:2014}, based on a 2$\sigma$ variation 
of the best fit values as of 2014.\\[-0.3in] 

\begin{figure}[floatfix, p!]
\begin{center}
  \includegraphics[width=5.5in,height=7.5in]{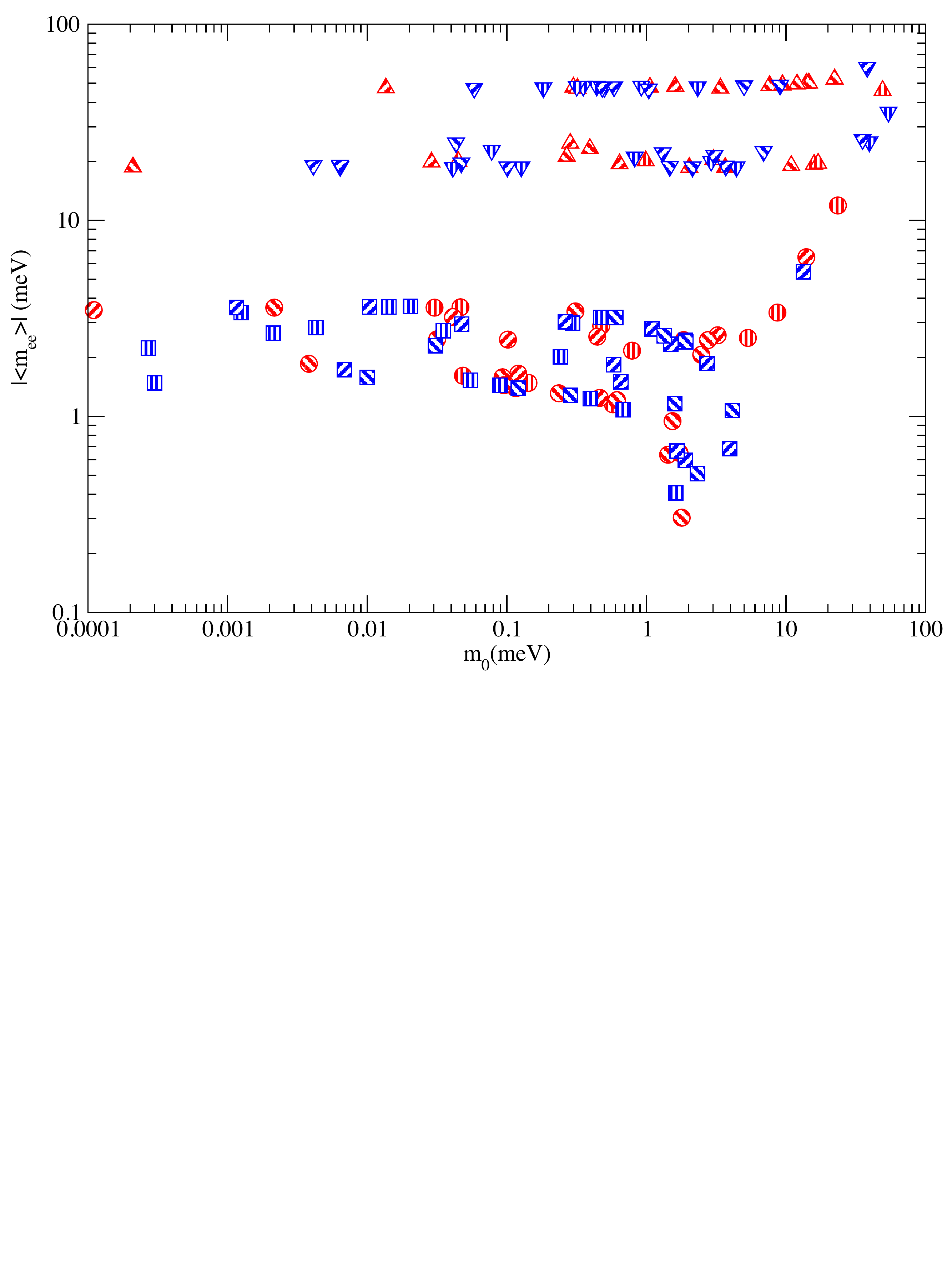}
  \vspace{-3.5in}
  \caption{Effective mass for neutrinoless double beta decay vs. the lightest
    neutrino mass for both NH and IH models.  The symbols for the models are
    the same as in Fig. 1.}
\label{fig:DBD}
\end{center}
\end{figure}

\section{Summary}
\vspace{-0.2in}
To explain quark and lepton masses and mixing angles, one has to extend the 
standard model by putting the quarks and leptons into irreducible 
representations of a discrete group. We argue that discrete flavor symmetries 
can be avoided, if we extend the gauge group to the point where the discrete 
symmetry is no longer needed. By consolidating flavor and family symmetries 
into a single gauged Lie group we eliminate the problems associated with 
discrete symmetry, e.g., violation by gravity, domain walls, etc.  We have 
given explicit examples of models having varying degrees of predictability 
obtained by scanning over groups and representations and identifying cases 
with operators contributing to mass and mixing matrices that need little 
fine-tuning of prefactors. Models in \SU{12} are particularly interesting. 

We have been guided by simplicity. Starting with SM$\times 
{\bf G}_{\rm flavor}$, 
we let SM$\rightarrow$ \SU{N} and increase $N$ until we can eliminate 
${\bf G}_{\rm flavor}$ and still fit known mass and mixing data. This process is 
rather involved. First we place the SM particles in \SU{5} irreps. Beginning 
with anomaly-free sets of irreps containing three families of fermions, we then 
assign the family $\mathbf{\bar{5}}$ and $\mathbf{10}$ irreps to \SU{N} irreps 
in a way that is consistent with known data. This requires scanning over 
fermion assignments and Higgs irreps to allow the necessary Yukawa 
coupling terms in the Largangian to generate successful models. The Higgs
irreps are also required to be capable of breaking the symmetry directly from 
\SU{N} to the SM without breaking SUSY.  Since there is an infinity of 
possible models, the scan is directed and limited in various ways toward 
finding the simplest class of examples.

We begin to find satisfactory models that require no discrete 
${\bf G}_{\rm flavor}$ 
symmetry at $N=12$. Smaller $N$ is insufficient to fit the data without 
keeping at least a small discrete flavor group. Larger $N$ typically gives too 
many parameters, hence we have focused on $N=12$, which seems to be 
the ``sweet spot'' for model building. In particular, the smallest anomaly-free 
set in \SU{12}  which is ${\bf 66} + {\bf 495} + 2({\bf \overline{220}}) + 
2({\bf \overline{12}})$ stands out for its simplicity. It contains six irreps, 
and it turns out that we can assign a single one of the six to each of the 
$\mathbf{\bar{5}}$'s and $\mathbf{10}$'s in the three families of \SU{5}. All 
other three-family sets in \SU{12} have more than six irreps, hence some of 
the irreps in the anomaly free sets cannot contain light fermions. We have  
limited our focus to this simplest anomaly-free set; however, there are still 
numerous issues to consider, e.g., which $\mathbf{\bar{5}}$ or 
$\mathbf{10}$ to assign to each of the \SU{12} irreps, which Higgs fields to 
include, how to include righthanded fermion singlet neutrinos, etc. To 
handle these issues we rely on scans over assignments. As described in 
the text, the scans systematically consider models, generate mass and 
mixing matrices, compare them with data, and those that do not drop by the 
wayside are kept, while those that fail tests along the way are eliminated 
from consideration. The result is approximately 30 models for each complete
scan (labeled $I$ and $II$ for their $\mathbf{10}$ assignments, and 
$A$, $B$, and $C$ for the 
assignment of their $\mathbf{\bar{5}}$'s) that satisfy our criteria of 
providing a fit to all known mass and mixing data which is little fine-tuned, 
while providing predictive power that can distinguish amongst our models and 
also discriminate between them and other  models in the literature. Once 
the fits of masses and mixings are complete for our models, they  allow us 
to make further predictions for the neutrino masses and hierarchy, the 
octant of the atmospheric mixing angle, leptonic CP violation, Majorana 
phases, and the effective mass observed in neutrinoless double beta decay.

Our purpose has been to unify family and flavor symmetries into a single 
gauge group.  What we have achieved is a demonstration that mass and 
mixing data can be fit within a class of models where the only symmetry is a 
gauged \SU{12}. Furthermore, these models can be predictive and 
distinguished from discrete flavor symmetry models. In addition,~$N = 12$ 
is small enough that it is conceivable that a model of this type can be 
contained within a compactification of the superstring. 

Among interesting features that have arisen in finding apparent satisfactory 
models are the following.  While we have emphasized the simplicity 
in assigning the three families of quarks and leptons to irreps of the 
smallest anomaly-free set of \SU{12}, the massive lefthanded conjugate 
(or righthanded) neutrinos must be placed in singlets of both \SU{5} and 
\SU{12}.  We have made the conventional choice of three such neutrinos, 
but it is clear that one could also have considered only two, or included  
several additional singlet sterile neutrinos in the model.

Two pairs of Higgs doublets appear in the models, 
(\irrep{5})\irrepbarsub{495}{H} and its conjugate along with 
(\irrepbar{5})\irrepbarsub{12}{H} and its conjugate, but only the first of each 
pair listed here are required to get EW VEVs in the successful models.  
The other two Higgs doublets, (\irrepbar{5})\irrepsub{495}{H} and 
(\irrep{5})\irrepsub{12}{H}, play no role in these models and can be 
considered inert.  Note the ``mismatch'' nature of the pair that develops 
VEVs and the other pair which does not.

In order to break the \SU{5} GUT symmetry, an adjunct \irrep{24} must be 
present in the model, but it can not be present in the \SU{12} adjunct 
\irrep{143} which would break SUSY at the \SU{12} scale.  Instead, two 
\irrep{24}'s emerge with the inclusion of \irrepsub{5148}{H} and 
\irrepbarsub{5148}{H} and their breakings at the \SU{12} scale.  Since they 
originate from an \SU{12} complex pair, a ready means arises of introducing 
CP phases in the models.  At the same time, their VEVs serve to split the 
spectra of the down quarks and charged leptons.

As for the model predictions for the unknown masses and mixing in the neutrino
sector, the successful models favor NH over IH by a ratio of 76 to 64 for the 
five runs considered..  The NH models
slightly favor the second octant for $\sin^2 \theta_{23}$, while the IH models 
are impartial to the first and second octants.  Many models favor small leptonic
CP violation, while a few favor larger violations.  Many pairs of successful 
models with identical Yukawa matrix textures have both NH and IH solutions 
due to different prefactors and $\Lambda_R$ scales emerging in the fitting 
procedure.  

Finally, we have shown for the successful models that the neutrinoless double
beta decay experiments may need to reach down to an effective mass 
$|\langle m_{ee}\rangle| \sim 10$ meV, in order to eliminate an inverted 
light neutrino mass hierarchy.  We also note that the present cosmological
constraints on the sum of the light neutrino masses \cite{Huang:2015}, 
$\Sigma m_{\nu,{\rm IH}} < 0.20$ eV, are insufficient to eliminate any of the 
apparently successful IH models.  
%\newpage
\begin{center}
{\bf ACKNOWLEDGMENTS}
\end{center}

  One of us (CHA) thanks the Fermilab Theoretical Division for its kind 
hospitality where his research was carried out.  The work of RPF was supported
by a fellowship within the Postdoc-Programme of the German Academic Exchange
Sevice (DAAD) and the Bundesministerium f\"{u}r Bildung und Forschung (BMBF)
under contract no. 05H12WWE.  The work of TWK was supported  by US DoE grant 
DE-FG05-85ER40226.  Fermilab is operated by Fermi Research Alliance, LLC under 
Contract No. De-AC02-07CH11359 with the U.S. Department of Energy.
\newpage
\begin{center}
{\bf APPENDIX A. Direct Breaking of \SU{12} $\rightarrow$ \SU{5}}
\end{center}

Complex irreps of \SU{N} all have non vanishing charges in the Cartan 
subalgebra.  If we give them VEVs, they then break a portion of that 
subalgebra. This in turn lowers the overall rank of the remaining symmetry 
group. Our interest here is in giving VEVs to antisymmetric irreps of 
\SU{12} to break the gauge symmetry directly to \SU{5}. These irreps are all 
complex except for the antisymmetric tensor with 6 indices which is real. A 
scheme for giving VEVs to antisymmetric tensor irreps of \SU{N} to reduce rank 
was devised in \cite{Frampton:1981pf,Frampton:1982mj} where vanishing total 
Dynkin weights for the VEVs provides gauge spontaneous symmetry breaking (SSB) 
without SUSY breaking.  

Here we demonstrate that the direct gauge SSB of \SU{12} $\rightarrow$ \SU{5}  
is possible without breaking SUSY by providing two solutions, one for the 
fourth anomaly-free set of Eq. \eqref{eq:afsets}, and the other for the 
simplest anomaly-free set of Eq. \eqref{eq:afsets}. 

The first example uses just the supermultiplets of the model where the chiral 
families live.  There the irreps we have to work with are
${\bf 66} + 2({\bf 495}) + {\bf \overline{792}} + 2({\bf \overline{220}}) + 
8({\bf \overline{12}}).$
Let us write VEVs with upper indices for the unbarred irreps, e.g.,
$v^{a,b}$ for the ${\bf 66}$, and with lower indices for the barred irreps, 
e.g., $v_{a,b,c,d,e}$ for the ${\bf \overline{792}}$. (We could instead
use an epsilon symbol with 12 indices to write the ${\bf \overline{792}}$ 
with 7 upper indices and likewise for other barred irreps, but it will 
not be necessary here.)

One set of chiral superpartner VEVs that breaks \SU{12} directly to \SU{5} is
\begin{equation}
v_{12,11,10,9,8},~~ v^{12,11,10,9},~~ v^{9,8,7,6}, ~~v_{7}, ~~v_{6},  ~~v_{9},
\end{equation}
where the VEVs can be in three different ${\bf \overline{12}}$s and in the two 
different ${\bf 495}$s.  We take all VEVs to be of equal magnitude.
The corresponding Dynkin weights given in the same order as the above VEVs 
are:\\[-0.1in]

\begin{equation}
\hspace{0.0in}\begin{tabular}{p{0.20in}p{0.20in}p{0.20in}p{0.20in}p{0.20in}p{0.20in}p{0.20in}p{0.20in}p{0.20in}p{0.20in}p{0.20in}p{0.20in}p{0.20in}}
 [&  1&  0&  0&  0&  0& -1&  0&  0&  0&  0&  0& ]\\

 [& -1&  0&  0&  0&  1&  0&  0&  0&  0&  0&  0& ]\\

 [&  0&  0&  0& -1&  0&  0&  0&  1&  0&  0&  0& ]\\

 [&  0&  0&  0&  0&  0&  1& -1&  0&  0&  0&  0& ]\\

 [&  0&  0&  0&  0&  0&  0&  1& -1&  0&  0&  0& ]\\

 [&  0&  0&  0&  1& -1&  0&  0&  0&  0&  0&  0& ]\\[0.25in]
\end{tabular}
\end{equation}

\noindent Summing the weights as in vector addition, we get zero total 
weight so SUSY remains unbroken.  Note that these VEVs give a minimum for the 
Higgs potential at zero, as required by SUSY. There could be some flat 
directions at this minimum, but since there are many terms in the 
superpotential this is not the generic situation.

The second example involves the simplest anomaly-free set which is of most 
interest in this paper, 
$  {\bf 66} + {\bf 495} + 2({\bf \overline{220}}) + 2({\bf \overline{12}}),$
with its scalar superpartners which are assumed to get VEVs, aside from the 
two ${\bf \overline{12}}$'s, along with a pair of Higgs singlets $\higgs{12}$ 
and $\higgsbar{12}$.  With the same tensor notation as above, we can 
form the following tensor contraction of the VEVS,
\begin{equation} 
v^{12,11,10,9}\ v_{12,11,10}\ v_{9,8,7}\ v^8\ v^{7,6}\ v_6.
\end{equation}
Again with all VEVs equal in magnitude, the ordered Dynkin weights are:
\\[-0.1in]
\begin{equation}
\hspace{0.0in}\begin{tabular}{p{0.20in}p{0.20in}p{0.20in}p{0.20in}p{0.20in}p{0.20in}p{0.20in}p{0.20in}p{0.20in}p{0.20in}p{0.20in}p{0.20in}p{0.20in}}
 [&  1&  0&  0&  0& -1&  0&  0&  0&  0&  0&  0& ]\\

 [& -1&  0&  0&  1&  0&  0&  0&  0&  0&  0&  0& ]\\

 [&  0&  0&  0& -1&  0&  0&  1&  0&  0&  0&  0& ]\\

 [&  0&  0&  0&  0&  1& -1&  0&  0&  0&  0&  0& ]\\

 [&  0&  0&  0&  0&  0&  1&  0& -1&  0&  0&  0& ]\\

 [&  0&  0&  0&  0&  0&  0& -1&  1&  0&  0&  0& ]\\[0.25in]
\end{tabular}
\end{equation}

\noindent The sum of the Dynkin weights vanishes, so $\SU{12} \rightarrow 
\SU{5}$ and SUSY remains unbroken. 

\begin{center} 
{\bf APPENDIX B.  Matrix Element Contributions to the Selected Model}
\end{center}

Here we present in Table IV the leading diagrams contributing to the 
Yukawa matrix elements for the quark and lepton mass matrices of the 
special model considered in Sect. IV C.  Several 
diagrams of the same dimension contribute to a given matrix element in 
many cases as listed.  These diagrams apply for the {\bf IC}6 class of 
models listed in Table II.

\input{matrixelementtable10C.tex}

\input{matrixelementtablecont10C.tex}

\end{document}

%% file: ModelClass10C.tex
\begin{table}
\begin{tabular}{lccccccccccc}
%\begin{tabularx}{\textwidth}{lX}\\[-9pt]
\hline\hline
\multicolumn{10}{c}{\bf Family Assignments:\qquad  I. (10):  \irrep{495_1}, 
\irrep{\overline{220}_2}, \irrep{66_3} \qquad II. (10):  
\irrep{\overline{220}_1}, \irrep{495_2}, \irrep{66_3}}\\
\multicolumn{10}{c}{\bf A.  $({\bf \overline{5}})$:  \irrep{\overline{12}_1}, 
  \irrep{\overline{12}_2}, \irrep{\overline{220}_3}\qquad 
 B. $({\bf \overline{5}})$:  \irrep{\overline{12}_1}, \irrep{\overline{220}_2}, 
\irrep{\overline{12}_3} \qquad 
C.  $({\bf \overline{5}})$:  \irrep{\overline{220}_1}, 
  \irrep{\overline{12}_2}, \irrep{\overline{12}_3}}\\
%\hline
 & {\bf Run} & \ {\bf MF1}  & {\bf MF2} & {\bf MF3} & {\bf MF4} & {\bf MF5} & 
  ({\bf 1}){\bf Higgs} 
 & \boldmath$\epsilon$  & \boldmath$\kappa$  & \boldmath$\arg(\kappa)$
   \\[0.05in]
\hline\hline
{\bf IA}1: & 1 & \irrep{66} & \irrep{220} & \irrep{495} &  &            
          & \hspace{0.3in} \irrep{66}, \irrep{792} \hspace{0.3in}  
          & -0.0117 \qquad  &  0.0436 & $35.8^\circ$ \\
{\bf IA}1: & 2 & \irrep{66} & \irrep{220} & \irrep{495} &  &            
          & \hspace{0.3in} \irrep{792} \hspace{0.3in}  
          & -0.0470 \qquad  &  0.0241 & $-119^\circ$ \\
{\bf IA}1: & 5 & \irrep{66} & \irrep{220} & \irrep{495} &  &            
          & \hspace{0.3in} \irrep{66}, \irrep{792} \hspace{0.3in}  
          & -0.0062 \qquad  &  0.0464 & $-164^\circ$ \\
{\bf IA}2: & 1 & \irrep{66} & \irrep{220} & \irrep{495} & \irrep{792} &        
          & \hspace{0.3in} \irrep{66}, \irrep{792} \hspace{0.3in}  
          & -0.0117 \qquad  &  0.0436 & $35.8^\circ$ \\
{\bf IA}2: & 2 &\irrep{66} & \irrep{220} & \irrep{495} & \irrep{792} &         
          & \hspace{0.3in} \irrep{792} \hspace{0.3in}  
          & -0.0470 \qquad  &  0.0241 & $-119^\circ$ \\
{\bf IA}2: & 5 & \irrep{66} & \irrep{220} & \irrep{495} & \irrep{792} &        
          & \hspace{0.3in} \irrep{66}, \irrep{792} \hspace{0.3in}  
          & -0.0062 \qquad  &  0.0464 & $-164^\circ$ \\
{\bf IA}3: & 1 & \irrep{12} & \irrep{66}  & \irrep{495} &  & 
          & \irrep{66}, \irrep{792}  &  0.0042  &  0.0268  &  $76.3^\circ$ \\ 
{\bf IA}3: & 3 & \irrep{12} & \irrep{66}  & \irrep{495} &  & 
          & \irrep{66}, \irrep{792}  &  -0.0111  &  0.0070  &  $43.1^\circ$ \\ 
{\bf IA}4: & 3 & \irrep{12} & \irrep{66}  & \irrep{495} & \irrep{792} & 
          & \irrep{66}, \irrep{792}  &  -0.0111  &  0.0070  &  $43.1^\circ$ \\ 
{\bf IA}4: & 4 & \irrep{12} & \irrep{66}  & \irrep{495} & \irrep{792} & 
          & \irrep{66}, \irrep{792}  &  -0.0077  &  0.0383  &  $-15.0^\circ$ \\ 
{\bf IA}5: & 3 & \irrep{12} & \irrep{66}  & \irrep{220} & \irrep{495} & 
          & \irrep{66}, \irrep{792}  &  0.0177  &  0.0742  &  $-171^\circ$ \\ 
{\bf IA}5: & 5 & \irrep{12} & \irrep{66}  & \irrep{220} & \irrep{495} & 
          & \irrep{66}, \irrep{792}  &  0.0114  &  0.0170  &  $-112^\circ$ \\ 
{\bf IA}6: & 1 & \irrep{12} & \irrep{66} & \irrep{220} & \irrep{495} & 
          \irrep{792} 
          & \irrep{66}, \irrep{792}  &  -0.0117  &  0.0061  &  $40.1^\circ$ \\ 
{\bf IA}6: & 3 & \irrep{12} & \irrep{66}  & \irrep{220} & \irrep{495} & 
          \irrep{792}
          & \irrep{66}, \irrep{792}  &  0.0177  &  0.0742  &  $-171^\circ$ \\ 
{\bf IA}6: & 5 & \irrep{12} & \irrep{66}  & \irrep{220} & \irrep{495} & 
          \irrep{792}
          & \irrep{66}, \irrep{792}  &  0.0114  &  0.0170  &  $-112^\circ$ \\ 
\hline
{\bf IB}1: & 4 & \irrep{66}  & \irrep{220} & \irrep{495} &        &         
          &  \irrep{66}, \irrep{792} & -0.0051  & 0.0245  &  $-128^\circ$ \\
{\bf IB}1: & 5 & \irrep{66}  & \irrep{220} & \irrep{495} &     &         
          &  \irrep{66}, \irrep{792} &  0.0116  & 0.0155  &  $56.6^\circ$ \\
{\bf IB}2: & 4 & \irrep{66}  & \irrep{220} & \irrep{495} & \irrep{792} &       
          &  \irrep{66}, \irrep{792} & -0.0051  & 0.0245  &  $-128^\circ$ \\
{\bf IB}2: & 5 & \irrep{66}  & \irrep{220} & \irrep{495} & \irrep{792} &      
          &  \irrep{66}, \irrep{792} &  0.0116  & 0.0155  &  $56.6^\circ$ \\
{\bf IB}3: & 4 & \irrep{12}  & \irrep{66} & \irrep{495} &        &         
          &  \irrep{66}, \irrep{792} & -0.0087  & 0.0060  &  $119^\circ$ \\
{\bf IB}4: & 1 & \irrep{12}  & \irrep{66} & \irrep{495} & \irrep{792} &         
          &  \irrep{66}, \irrep{792} & -0.0094  & 0.0230  &  $121^\circ$ \\
{\bf IB}4: & 4 & \irrep{12}  & \irrep{66} & \irrep{495} & \irrep{792} &         
          &  \irrep{66}, \irrep{792} & -0.0087  & 0.0060  &  $119^\circ$ \\
{\bf IB}5: & 5 & \irrep{12}  & \irrep{66} & \irrep{220} & \irrep{495} &         
          &  \irrep{66}, \irrep{792} & -0.0125  & 0.0124  &  $121^\circ$ \\
{\bf IB}6: & 3 & \irrep{12}  & \irrep{66} & \irrep{220} & \irrep{495} & 
          \irrep{792}
          &  \irrep{792} & -0.0244  &  0.1430  &  $10.2^\circ$ \\
{\bf IB}6: & 5 & \irrep{12}  & \irrep{66} & \irrep{220} & \irrep{495} & 
          \irrep{792}
          &  \irrep{792} & -0.0125  &  0.0124  &  $121^\circ$ \\
\hline
\end{tabular}
\end{table}

%\newpage
\begin{table}
\begin{tabular}{lccccccccccccc}
%\begin{tabularx}{\textwidth}{lX}\\[-9pt]
 & {\bf Run} & \ {\bf MF1}  & {\bf MF2} & {\bf MF3} & {\bf MF4} & {\bf MF5}   
  & ({\bf 1}){\bf Higgs} & \boldmath$\epsilon$ 
 &   & \boldmath$\kappa$ &   & \boldmath$\arg(\kappa)$
   \\[0.05in]
\hline\hline
{\bf IC}1: & 1 & \irrep{66} &  \irrep{220} & \irrep{495} &         &
          & \irrep{66}, \irrep{792} & 0.0174 & \hspace{0.3in} & 0.0262 & \hspace{0.3in}
 & $-130^\circ$ \\[-0.05in]
{\bf IC}1: & 4 & \irrep{66} & \irrep{220} & \irrep{495} &         &
          & \irrep{66}, \irrep{792} & 0.0064 & & 0.0272 &  & $-97.1^\circ$ \\[-0.05in]
{\bf IC}1: & 5 & \irrep{66} & \irrep{220} & \irrep{495} &         &
          &  \irrep{66}, \irrep{792} & -0.0125 &  & 0.0080 & & $52.8^\circ$ \\[-0.05in]
{\bf IC}2: & 1 & \irrep{66} & \irrep{220} & \irrep{495} & \irrep{792} & 
          & \irrep{66}, \irrep{792}  &  0.0174 & & 0.0263 &  & $-130^\circ$ \\[-0.05in]
{\bf IC}2: & 4 & \irrep{66} &   \irrep{220} & \irrep{495} & \irrep{792} & 
          & \hspace{0.3in} \irrep{66}, \irrep{792} \hspace{0.3in}  &  0.0064  
          &  & 0.0272 &  &  $-97.1^\circ$ \\[-0.05in]
{\bf IC}2: & 5 & \irrep{66} &   \irrep{220} & \irrep{495} & \irrep{792} & 
          & \irrep{66},  \irrep{792} & -0.0125 &  & 0.0080 &  & $52.8^\circ$ \\[-0.05in]
{\bf IC}3: & 1 & \irrep{12} &   \irrep{66} & \irrep{495} &             &  
          &  \irrep{66}, \irrep{792} &  -0.0088 &  & 0.0381 &  & $13.4^\circ$ \\[-0.05in]
{\bf IC}3: & 2 & \irrep{12} &   \irrep{66} & \irrep{495} &             & 
          &  \irrep{66}, \irrep{792} &  -0.0066 &  & 0.0327 &  & $29.6^\circ$ \\[-0.05in]
{\bf IC}3: & 3 & \irrep{12} &  \irrep{66} & \irrep{495} &             &  
          &  \irrep{66}, \irrep{792} &  -0.0079 &  & 0.0317 &  & $18.8^\circ$ \\[-0.05in]
{\bf IC}3: & 5 & \irrep{12} &   \irrep{66} & \irrep{495} &         &
          & \irrep{66},  \irrep{792} & -0.0105 &  & 0.0035 &  & $61.8^\circ$ \\[-0.05in]
{\bf IC}4: & 1 & \irrep{12} &  \irrep{66}  & \irrep{495}  & \irrep{792} & 
          & \irrep{66}, \irrep{792}  & -0.0088 &  & 0.0381 &  & $13.4^\circ$ \\[-0.05in]
{\bf IC}4: & 2 & \irrep{12} &  \irrep{66}  & \irrep{495}  & \irrep{792} &
          & \irrep{66}, \irrep{792}  & -0.0066  &  & 0.0327 &  & $29.6^\circ$ \\[-0.05in]
{\bf IC}4: & 3 & \irrep{12} &  \irrep{66}  & \irrep{495}  & \irrep{792} & 
          & \irrep{66}, \irrep{792}  & -0.0079 &  & 0.0317 &  & $18.9^\circ$ \\[-0.05in]
{\bf IC}4: & 5 & \irrep{12} &  \irrep{66} & \irrep{495} & \irrep{792} &  
          & \irrep{66}, \irrep{792}  &  -0.0105 &  & 0.0035 & & $61.8^\circ$ \\[-0.05in]
{\bf IC}5: & 1 & \irrep{12} &  \irrep{66} & \irrep{220} & \irrep{495} &  
          & \irrep{66}, \irrep{792}  & -0.0149 & & 0.0357 &  & $16.7^\circ$\\[-0.05in]
{\bf IC}5: & 4 & \irrep{12} &  \irrep{66} & \irrep{220} & \irrep{495} &  
          & \irrep{66}, \irrep{792}  &  0.0145 &  & 0.0231 &  & $27.5^\circ$\\[-0.05in]
{\bf IC}6: & 1 & \irrep{12} &  \irrep{66} & \irrep{220} & \irrep{495}  
           & \irrep{792}\qquad & \irrep{66}, \irrep{792}  
          & -0.0149 & & 0.0357 & & $16.8^\circ$ \\[-0.05in]
{\bf IC}6: & 4 & \irrep{12} &  \irrep{66} & \irrep{220} & \irrep{495} 
          & \irrep{792} & \irrep{66}, \irrep{792}  
          & 0.0145 & & 0.0231 & & $27.5^\circ$ \\[-0.05in]
\hline
{\bf IIA}1: & 2 & \irrep{66} & \irrep{220} & \irrep{495} &            &
          & \irrep{66}, \irrep{792}  & -0.0026 & & 0.0138 & & $77.2^\circ$ \\[-0.05in]
{\bf IIA}1: & 4 & \irrep{66} & \irrep{220} & \irrep{495} &            &
          & \irrep{66}, \irrep{792}  & -0.0103 & & 0.0154 & & $58.4^\circ$ \\[-0.05in]
{\bf IIA}2: & 2 & \irrep{66} & \irrep{220} & \irrep{495} & \irrep{792}      &
          & \irrep{66}, \irrep{792}  & -0.0026 & & 0.0138 & & $77.2^\circ$ \\[-0.05in]
{\bf IIA}2: & 4 & \irrep{66} & \irrep{220} & \irrep{495} & \irrep{792}      &
          & \irrep{66}, \irrep{792}  & -0.0103 & & 0.0154 & & $58.4^\circ$ \\[-0.05in]
{\bf IIA}3: & 1 & \irrep{12} & \irrep{66} & \irrep{495} &            &
          & \irrep{66}, \irrep{792}  & -0.0150 & & 0.0337 & & $-9.1^\circ$ \\[-0.05in]
{\bf IIA}3: & 2 & \irrep{12} & \irrep{66} & \irrep{495} &            &
          & \irrep{66}, \irrep{792}  & -0.0139 & & 0.0087 & & $60.8^\circ$ \\[-0.05in]
{\bf IIA}3: & 3 & \irrep{12} & \irrep{66} & \irrep{495} &            &
          & \irrep{66}, \irrep{792}  & 0.0119 & & 0.0396 & & $170^\circ$ \\[-0.05in]
{\bf IIA}3: & 5 & \irrep{12} & \irrep{66} & \irrep{495} &            &
          & \irrep{66}, \irrep{792}  & 0.0157 & & 0.0312 & & $-132^\circ$ \\[-0.05in]
{\bf IIA}4: & 1 & \irrep{12} & \irrep{66} & \irrep{495} & \irrep{792} & 
          & \irrep{66}, \irrep{792}  & -0.0150 & & 0.0337 & & $-9.1^\circ$ \\[-0.05in] 
{\bf IIA}4: & 2 & \irrep{12} & \irrep{66} & \irrep{495} & \irrep{792} & 
          & \irrep{66}, \irrep{792}  & -0.0139 & & 0.0087 & & $60.8^\circ$ \\[-0.05in] 
{\bf IIA}4: & 3 & \irrep{12} & \irrep{66} & \irrep{495} & \irrep{792} & 
          & \irrep{66}, \irrep{792}  & 0.0158 & & 0.0274 & & $84.6^\circ$ \\[-0.05in] 
{\bf IIA}4: & 5 & \irrep{12} & \irrep{66} & \irrep{495} & \irrep{792} & 
          & \irrep{66}, \irrep{792}  & 0.0119 & & 0.0396 & & $170^\circ$ \\[-0.05in] 
{\bf IIA}5: & 1 & \irrep{12} & \irrep{66} & \irrep{220} & \irrep{495} & 
          & \irrep{66}, \irrep{792} & -0.0113 & & 0.0334 & & $9.3^\circ$ \\[-0.05in] 
{\bf IIA}5: & 3 & \irrep{12} & \irrep{66} & \irrep{220} & \irrep{495} & 
          & \irrep{66}, \irrep{792}  & -0.0125 & & 0.0080 & & $79.7^\circ$ \\[-0.05in] 
{\bf IIA}6: & 1 & \irrep{12} & \irrep{66} & \irrep{220} & \irrep{495} 
          & \irrep{792} & \irrep{66}, \irrep{792} 
          & -0.0113 & & 0.0334 & & $9.3^\circ$ \\[-0.05in] 
{\bf IIA}6: & 2 & \irrep{12} & \irrep{66} & \irrep{220} & \irrep{495} 
          & \irrep{792} & \irrep{66}, \irrep{792} 
          & 0.0079 & & 0.0129 & & $37.7^\circ$ \\[-0.05in] 
{\bf IIA}6: & 3 & \irrep{12} & \irrep{66} & \irrep{220} & \irrep{495} 
          & \irrep{792} & \irrep{66}, \irrep{792} 
          & -0.0125 & & 0.0080 & & $79.7^\circ$ \\[-0.05in] 
{\bf IIA}6: & 4 & \irrep{12} & \irrep{66} & \irrep{220} & \irrep{495} 
          & \irrep{792} & \irrep{66}, \irrep{792} 
          & 0.0067 & & 0.0225 & & $-124^\circ$ \\[-0.05in] 
\hline
\end{tabular}
\end{table}
%\newpage
\begin{table}
\begin{tabular}{lccccccccccccc}
%\begin{tabularx}{\textwidth}{lX}\\[-9pt]
 & {\bf Run} & \ {\bf MF1}  & {\bf MF2} & {\bf MF3} & {\bf MF4} & {\bf MF5}   
  & ({\bf 1}){\bf Higgs} & \boldmath$\epsilon$ 
 &   & \boldmath$\kappa$ &   & \boldmath$\arg(\kappa)$
   \\[0.05in]
\hline\hline
{\bf IIB}1: & 3 & \irrep{66}  & \irrep{220} & \irrep{495} &           &         
          &  \irrep{66}, \irrep{792} & 0.0074 & \hspace{0.3in} & 0.0719 & \hspace{0.3in} & $52.3^\circ$ \\[-0.05in]
{\bf IIB}2: & 3 & \irrep{66} & \irrep{220} & \irrep{495} & \irrep{792} & 
              &  \irrep{792} & 0.0318 & & 0.0282 & & $65.7^\circ$ \\[-0.05in]
{\bf IIB}2: & 3 & \irrep{66}  & \irrep{220} & \irrep{495} & \irrep{792}   &    
          &  \irrep{66}, \irrep{792} & 0.074 & & 0.0719 & & $52.3^\circ$ \\[-0.05in]
{\bf IIB}2: & 5 & \irrep{66}  & \irrep{220} & \irrep{495} & \irrep{792}   &    
          &  \irrep{66}, \irrep{792} & 0.0071 & & 0.0221 & & $34.4^\circ$ \\[-0.05in]
{\bf IIB}3: & 1 & \irrep{12}  & \irrep{66} & \irrep{495} &           &         
          &  \irrep{66}, \irrep{792} & 0.0102 & & 0.0202 & & $36.1^\circ$ \\[-0.05in]
{\bf IIB}3: & 2 & \irrep{12}  & \irrep{66} & \irrep{495} &           &         
          &  \irrep{66}, \irrep{792} & 0.0219 & & 0.0253 & & $34.4^\circ$ \\[-0.05in]
{\bf IIB}3: & 3 & \irrep{12}  & \irrep{66} & \irrep{495} &           &         
          &  \irrep{66}, \irrep{792} & -0.0131 & & 0.0077 & & $37.1^\circ$ \\[-0.05in]
{\bf IIB}3: & 4 & \irrep{12}  & \irrep{66} & \irrep{495} &           &         
          &  \irrep{66}, \irrep{792} & -0.0100 & & 0.0276 & & $27.2^\circ$ \\[-0.05in]
{\bf IIB}4: & 1 & \irrep{12} & \irrep{66} & \irrep{495} & \irrep{792} &         
          & \hspace{0.3in} \irrep{66}, \irrep{792} \hspace{0.3in} & 0.0102 & & 0.0202 & & $36.1^\circ$ \\[-0.05in]
{\bf IIB}4: & 2 & \irrep{12} & \irrep{66} & \irrep{495} & \irrep{792} &         
          & \irrep{66}, \irrep{792}   & 0.0219 & & 0.0253 & & $34.4^\circ$ \\[-0.05in]
{\bf IIB}4: & 3 & \irrep{12} & \irrep{66} & \irrep{495} & \irrep{792} &         
          & \irrep{66}, \irrep{792}   & -0.0100 & & 0.0276 & & $27.2^\circ$ \\[-0.05in]
{\bf IIB}4: & 4 & \irrep{12} & \irrep{66} & \irrep{495} & \irrep{792} &         
          & \irrep{66}, \irrep{792}   & -0.015 & & 0.0387 & & $14.4^\circ$ \\[-0.05in]
{\bf IIB}5: & 2 & \irrep{12}  & \irrep{66} & \irrep{220} & \irrep{495} &       
          &  \irrep{66}, \irrep{792} & -0.0080 &  & 0.0254 & & $-122^\circ$ \\[-0.05in]
{\bf IIB}5: & 3 & \irrep{12}  & \irrep{66} & \irrep{220} & \irrep{495} &       
          &  \irrep{66}, \irrep{792} & 0.0147 & & 0.0115 & & $-146^\circ$ \\[-0.05in]
{\bf IIB}5: & 4 & \irrep{12} & \irrep{66} & \irrep{220} & \irrep{495} &         
          & \irrep{66}, \irrep{792}   & -0.037 & & 0.0232 & & $87.4^\circ$ \\[-0.05in]
{\bf IIB}6: & 2 & \irrep{12}  & \irrep{66} & \irrep{220} & \irrep{495} 
          & \irrep{792} & \irrep{66}, \irrep{792}  
          & -0.0080 & & 0.0254 &  & $-122^\circ$ \\[-0.05in]
{\bf IIB}6: & 3 & \irrep{12}  & \irrep{66} & \irrep{220} & \irrep{495} 
          & \irrep{792} & \irrep{66}, \irrep{792}  
          & 0.0044 & & 0.0359 & & $91.0^\circ$ \\[-0.05in]
{\bf IIB}6: & 4 & \irrep{12}  & \irrep{66} & \irrep{220} & \irrep{495} 
          & \irrep{792} & \irrep{66}, \irrep{792}  
          & -0.0038 & & 0.0232 & & $87.5^\circ$ \\[-0.05in]
\hline
{\bf IIC}1: & 1 & \irrep{66} & \irrep{220} & \irrep{495} &           &         
          &  \irrep{66}, \irrep{792} & 0.0111 & & 0.0230 & & $63.3^\circ$ \\[-0.05in]
{\bf IIC}1: & 2 & \irrep{66} & \irrep{220} & \irrep{495} &           &         
          &  \irrep{66}, \irrep{792} & -0.0037 & & 0.0281 & & $28.6^\circ$ \\[-0.05in]
{\bf IIC}2: & 2 & \irrep{66} & \irrep{220} & \irrep{495} & \irrep{792}  &    
          &  \irrep{66}, \irrep{792} & -0.0037 & & 0.0254 & & $57.0^\circ$ \\[-0.05in]
{\bf IIC}2: & 5 & \irrep{66} & \irrep{220} & \irrep{495} & \irrep{792}  &    
          &  \irrep{66}, \irrep{792} & 0.0072 & & 0.0160 & & $-87.7^\circ$ \\[-0.05in]
{\bf IIC}3: & 2 & \irrep{12}  & \irrep{66} & \irrep{495} &       &         
          & \irrep{66}, \irrep{792}  & 0.0028 & & 0.0408 & & $80.3^\circ$ \\[-0.05in]
{\bf IIC}3: & 4 & \irrep{12}  & \irrep{66} & \irrep{495} &       &         
          & \irrep{66}, \irrep{792}  & -0.0053 & & 0.0138 & & $13.4^\circ$ \\[-0.05in]
{\bf IIC}4: & 2 & \irrep{12}  & \irrep{66} & \irrep{495} & \irrep{792} &        
          & \irrep{66}, \irrep{792}  & -0.0037 & & 0.0281 & & $28.6^\circ$\\[-0.05in]
{\bf IIC}4: & 4 & \irrep{12}  & \irrep{66} & \irrep{495} & \irrep{792} &        
          & \irrep{66}, \irrep{792}  & -0.0053  & & 0.0138 & & $13.4^\circ$\\[-0.05in]
{\bf IIC}5: & 1 & \irrep{12}  & \irrep{66} & \irrep{220} & \irrep{495} &        
          & \irrep{792}  & -0.0575 & & 0.0266 & & $20.9^\circ$\\[-0.05in]
{\bf IIC}5: & 4 & \irrep{12}  & \irrep{66} & \irrep{220} & \irrep{495} &        
          & \irrep{66}, \irrep{792}  & 0.0086 & & 0.0223 & & $169^\circ$\\[-0.05in]
{\bf IIC}6: & 1 & \irrep{12}  & \irrep{66} & \irrep{220} & \irrep{495} 
          & \irrep{792} & \irrep{792}  & -0.0575 & & 0.0266 & & $20.9^\circ$\\[-0.05in]
{\bf IIC}6: & 4 & \irrep{12}  & \irrep{66} & \irrep{220} & \irrep{495} 
          & \irrep{792} & \irrep{66}, \irrep{792} 
          & 0.0086 & & 0.0222 & & $169^\circ$\\
\hline\hline
\end{tabular}
\caption{\label{tab:SummarytableI}Summary of irreps and model parameters for 
successful SU(12) models found in the five search and fitting runs.  The 
classes of models are labeled by their SU(5) $({\bf 10})$ and 
$({\bf \overline{5}})$ family assignments, while the righthanded Majorana 
neutrinos all belong to \SU{5} and \SU{12} singlets.  It is to be understood 
that the massive fermions and Higgs singlets occur in both unbarred and barred 
irreps.} 
\end{table}

%% file: ModelNeutrino10C.tex
\begin{table}
\begin{tabular}{lccccccccccccc}
%\begin{tabularx}{\textwidth}{lX}\\[-9pt]
\hline\hline
 & {\bf MH} & $\mathbf{\Lambda_R}$ & $\mathbf{M_1}$ & $\mathbf{M_2}$ 
 & $\mathbf{M_3}$ & $\mathbf{m_1}$ & $\mathbf{m_2}$ & $\mathbf{m_3}$ 
 & \boldmath$\sin^2 \theta_{23}$ & \boldmath$\delta$
 & \boldmath$\phi_1$ & \boldmath$\phi_2$ & \boldmath$|\langle m_{ee}\rangle|$ 
    \\[-0.05in]
 &    	    & \multicolumn{4}{c}{-------- (${\bf 10^{14}\ GeV}$) ---------}   
 & \multicolumn{3}{c}{------ ({\bf meV})------} 
 &        &       &        &     &    ({\bf meV}) \\[0.05in]
 \hline\hline
 {\bf IA}4: & NH & 0.174  &  0.0569   & 0.189  &  0.388  &  0.116  &  8.6  
 & 49.9 &  0.387  &  $-1.3^\circ$  & $5.6^\circ$ & $185^\circ$ & 1.39 \\[-0.1in]
 	& IH  & 0.0358    & 0.0496  & 0.0583  & 0.112  &  49.3  & 50.0 & 3.67
 & 0.390 & $178^\circ$ & $-174^\circ$ & $6.4^\circ$ & 18.6 \\
\hline 
{\bf IB}1: & NH & 0.0499 & 0.0055 & 0.0371 & 0.101 & 1.85 & 8.79 & 49.9 
& 0.510 & $-10.6^\circ$ & $-132^\circ$ & $40.6^\circ$ & 2.45 \\[-0.1in]
{\bf IB}2: & IH & 0.0102 & 0.0093 & 0.0182 & 0.0217 & 49.2 & 49.9 & 1.61 
& 0.545 & $-12.1^\circ$ & $235^\circ$ & $-133^\circ$ & 48.2 \\[-0.1in]
{\bf IB}3: & NH & 0.0011 & 0.0002 & 0.0062 & 0.0187 & 14.0 & 16.5 & 51.9 
& 0.469 & $134^\circ$ & $167^\circ$ & $-345^\circ$ & 6.48 \\[-0.1in]
           & IH & 0.0048 & 0.0021 & 0.0094 & 0.0097 & 49.2 & 49.9 & 1.06 
& 0.647 & $-121^\circ$ & $-44.0^\circ$ & $-64.8^\circ$ & 47.7 \\
{\bf IB}4: & NH & 0.0152 & 0.0223 & 0.0484 & 0.0496 & 0.121 & 8.60 & 49.9 
& 0.443 & $17.3^\circ$ & $-79.8^\circ$ & $114^\circ$ & 1.65 \\[-0.1in]
           & IH & 0.0088 & 0.0086 & 0.0128 & 0.0196 & 49.2 & 49.9 & 0.642
& 0.426 & $-158^\circ$ & $279^\circ$ & $85.6^\circ$ & 19.4 \\
\hline
{\bf IC}1: & NH & 3.38 & 0.0142 & 0.0243 & 10.1 & 0.0022 & 8.6 & 49.9 
& 0.543 & $82.7^\circ$ & $185^\circ$ & $219^\circ$ & 3.58 \\[-0.1in]
           & IH & 0.0009 & 0.0009 & 0.0013 & 0.0059 & 49.2 & 49.9 & 0.394 
& 0.481 & $-62.6^\circ$ & $-263^\circ$ & $241^\circ$ & 23.2 \\[-0.1in]
{\bf IC}2: & NH & 0.0520 & 0.0542 & 0.111 & 0.145 & 1.73 & 8.77 & 49.9 
& 0.630 & $97.0^\circ$ & $-227^\circ$ & $341^\circ$ & 0.65 \\[-0.1in]
           & IH & 0.0187 & 0.0228 & 0.0336 & 0.0544 & 49.2 & 49.9 & 0.269 
& 0.658 & $81.0^\circ$ & $-163^\circ$ & $-316^\circ$ & 21.2 \\[-0.1in]
{\bf IC}5: & IH & 0.013 & 0.0074 & 0.242 & 0.0319 & 49.2 & 49.9 & 0.321 
& 0.560 & $-90.2^\circ$ & $200^\circ$ & $172^\circ$ & 47.1 \\[-0.1in]
{\bf IC}6: & NH & 0.0155 & 0.0057 & 0.0176 & 0.0435 & 2.76 & 9.03 & 50.0 
& 0.460 & $-121^\circ$ & $-219^\circ$ & $-67.4^\circ$ & 2.45 \\[-0.1in]
           & IH & 0.0127 & 0.0105 & 0.0226 & 0.0332 & 49.3 & 50.0 & 3.39
& 0.600 & $-50.2^\circ$ & $-319^\circ$ & $12.8^\circ$ & 47.2 \\
\hline
 {\bf IIA}1: & NH & 0.0320  &  0.0069 & 0.0515 & 0.0543 & 0.469 & 8.61 & 49.9 
 & 0.360 &  $22.8^\circ$  &  $70.8^\circ$ & $-86.0^\circ$ & 3.2 \\[-0.1in]
 	& IH  & 0.0078  & 0.0123  & 0.0159  & 0.0190  & 49.2  & 50.0 & 2.91 
 & 0.578 & $140^\circ$ & $-32.5^\circ$ & $166^\circ$ & 20.0 \\[-0.1in]
{\bf IIA}2: & NH & 0.313 & 0.0501 & 0.250 & 0.750 & 0.00027 & 8.6 & 49.9 
& 0.590 & $-152^\circ$ & $-103^\circ$ & $64.0^\circ$ & 2.23 \\[-0.1in]
{\bf IIA}6: & NH & 0.127 & 0.0307 & 0.140 & 0.295 & 0.0143 & 8.6 & 49.9 
& 0.348 & $11.1^\circ$ & $308^\circ$ & $317^\circ$ & 3.61 \\[-0.1in]
        & IH & 0.0252 & 0.0149 & 0.0383 & 0.0507 & 49.2 & 49.9 & 0.317 
& 0.657 & $-161^\circ$ & $-251^\circ$ & $102^\circ$ & 48.2 \\
\hline
{\bf IIB}3: & NH & 0.0458 & 0.0635 & 0.0956 & 0.107 & 1.91 & 8.81 & 49.9 
& 0.447 & $-12.4^\circ$ & $-121^\circ$ & $-313^\circ$ & 2.44 \\[-0.1in]
{\bf IIB}4: & NH & 0.225 & 0.0416 & 0.400 & 0.438 & 0.0104 & 8.6 & 49.9 
& 0.419 & $-9.3^\circ$ & $-135^\circ$ & $-325^\circ$ & 3.61 \\[-0.1in]
           & IH & 0.741 & 0.0532 & 0.138 & 1.24 & 49.2 & 49.9 & 0.0041
& 0.413 & $-167^\circ$ & $134^\circ$ & $304^\circ$ & 19.0 \\[-0.1in]
{\bf IIB}5: & NH & 0.0055 & 0.0051 & 0.0130 & 0.0157 & 13.3 & 15.8 & 51.7 
& 0.449 & $-7.4^\circ$ & $-324^\circ$ & $207^\circ$ & 5.46 \\[-0.1in]
           & IH & 0.0072 & 0.0066 & 0.0182 & 0.0243 & 49.3 & 50.0 & 3.7 
& 0.504 & $11.3^\circ$ & $310^\circ$ & $-224^\circ$ & 18.9 \\[-0.1in]
{\bf IIB}6: & NH & 0.0337 & 0.0261 & 0.0634 & 0.0671 & 2.72 & 9.02 & 50.0 
& 0.415 & $170^\circ$ & $-146^\circ$ & $42.3^\circ$ & 1.86 \\
\hline
{\bf IIC}3: & NH & 0.0071 & 0.0127 & 0.0158 & 0.0215 & 1.92 & 8.80 & 49.9 
& 0.639 & $177^\circ$ & $-168^\circ$ & $13.2^\circ$ & 2.40 \\[-0.1in]
{\bf IIC}4: & NH & 0.0200 & 0.0327. & 0.0580 & 0.0717 & 0.121 & 8.6 & 49.9 
& 0.519 & $-178^\circ$ & $353^\circ$ & $172^\circ$ & 1.39 \\[-0.1in]
{\bf IIC}5: & IH & 0.0085 & 0.0161 & 0.0323 & 0.0439 & 50.0 & 50.7 & 9.07
& 0.469 & $179^\circ$ & $-177^\circ$ & $-176^\circ$ & 48.9 \\[-0.1in]
{\bf IIC}6: & NH & 0.191 & 0.0419 & 0.216 & 0.508 & 0.286 & 8.6 & 49.9 
& 0.598 & $179^\circ$ & $2.7^\circ$ & $-177^\circ$ & 1.28 \\[-0.1in]
            & IH & 0.196 & 0.0352 & 0.0473 & 0.0492 & 49.4 & 50.2 & 5.01 
& 0.551 & $-179^\circ$ & $-182^\circ$ & $177^\circ$ & 48.5 \\
\hline\hline
\end{tabular}
\caption{\label{tab:SummarytableI}Summary of successful SU(12) neutrino models
for the 4th run with $\Lambda_R$ as an additional fit parameter. 
The models are grouped according to the family assignments listed in Table II. 
In many cases, both
normal hierarchy NH and inverted hierarchy IH models have been generated for 
the same class of
family assignment and matrix textures with different sets of matrix element 
prefactors and $\Lambda_R$.}
\end{table}

%% file: matrixelementtable10C.tex
\begin{table}
\renewcommand\boldirrep\relax
\setlength{\arraycolsep}{1pt}
\begin{center}
\begin{tabular}{lll}\\[-14pt]
\hline\hline
\multicolumn{3}{c}{{\bf Matrix Element Contributions for Model IC}6}\\
\multicolumn{3}{l}{{\bf Fermions:}\hspace{0.61in} (\irrep{10})\irrepsub{495}{1}, 
    (\irrep{10})\irrepbarsub{220}{2}, (\irrep{10})\irrepsub{66}{3}, 
    (\irrepbar{5})\irrepbarsub{220}{1}, (\irrepbar{5})\irrepbarsub{12}{2}, 
    (\irrepbar{5})\irrepbarsub{12}{3}}\\[-0.1in]
\multicolumn{3}{l}{{\bf Massive Fermions:} \irrep{12}, \irrepbar{12}, \irrep{66}, \irrepbar{66}, 
    \irrep{220}, \irrepbar{220}, \irrep{495}, \irrepbar{495}, 
    \irrep{792}, \irrepbar{792}}\\[-0.1in]
\multicolumn{3}{l}{{\bf Higgs:}\hspace{0.6in}(\irrep{5})\irrepbarsub{495}{H}, (\irrepbar{5})		        \irrepbarsub{12}{H},
    (24)\irrepsub{5148}{H}, (24)\irrepbarsub{5148}{H}, (1)\irrepsub{1}{H}, (1)\irrepsub{66}{H},
    (1)\irrepbarsub{66}{H}, (1)\irrepsub{792}{H}, (1)\irrepbarsub{792}{H}}\\
\hline
\multicolumn{3}{l}{\bf Leading Up-Type Diagrams:}\\[-0.1in]
\textbf{Dim 4:} \\[-0.1in]
&    \UpType{3}{3} & (\irrep{10})\irrepsub{66}{3}.(\irrep{5})\irrepbarsub{495}{H}.(\irrep{10})\irrepsub{66}{3}\\[-0.1in]
\textbf{Dim 5:} \\[-0.1in]
&    \UpType{1}{3} & (\irrep{10})\irrepsub{495}{1}.(1)\irrepbarsub{66}{H}.\massivefermionpair{(\irrepbar{10})\irrepbar{66}}{(\irrep{10})\irrep{66}}.(\irrep{5})\irrepbarsub{495}{H}.(\irrep{10})\irrepsub{66}{3}\\[-0.1in]
&    \UpType{3}{1} & (\irrep{10})\irrepsub{66}{3}.(\irrep{5})\irrepbarsub{495}{H}.\massivefermionpair{(\irrep{10})\irrep{66}}{(\irrepbar{10})\irrepbar{66}}.(1)\irrepbarsub{66}{H}.(\irrep{10})\irrepsub{495}{1}\\[-0.1in]
&    \UpType{2}{3} & (\irrep{10})\irrepbarsub{220}{2}.(1)\irrepsub{792}{H}.\massivefermionpair{(\irrepbar{10})\irrepbar{66}}{(\irrep{10})\irrep{66}}.(\irrep{5})\irrepbarsub{495}{H}.(\irrep{10})\irrepsub{66}{3}\\[-0.1in]
&    \UpType{2}{3} & (\irrep{10})\irrepbarsub{220}{2}.(24)\irrepsub{5148}{H}.\massivefermionpair{(\irrepbar{10})\irrepbar{66}}{(\irrep{10})\irrep{66}}.(\irrep{5})\irrepbarsub{495}{H}.(\irrep{10})\irrepsub{66}{3}\\[-0.1in]
&    \UpType{3}{2} & (\irrep{10})\irrep{66_3}.(\irrep{5})\irrepbarsub{495}{H}.\massivefermionpair{(\irrep{10})\irrep{66}}{(\irrepbar{10})\irrepbar{66}}.(1)\irrepsub{792}{H}.(\irrep{10})\irrepbarsub{220}{2}\\[-0.1in]
&    \UpType{3}{2} & (\irrep{10})\irrep{66_3}.(\irrep{5})\irrepbarsub{495}{H}.\massivefermionpair{(\irrep{10})\irrep{66}}{(\irrepbar{10})\irrepbar{66}}.(24)\irrepsub{5148}{H}.(\irrep{10})\irrepbarsub{220}{2}\\[-0.1in]
\textbf{Dim 6:}\\[-0.1in]
&    \UpType{1}{1} & (\irrep{10})\irrepsub{495}{1}.(1)\irrepbarsub{66}{H}.\massivefermionpair{(\irrepbar{10})\irrepbar{66}}{(\irrep{10})\irrep{66}}.(\irrep{5})\irrepbarsub{495}{H}.\massivefermionpair{(\irrep{10})\irrep{66}}{(\irrepbar{10})\irrepbar{66}}.(1)\irrepbarsub{66}{H}.(\irrep{10})\irrepsub{495}{1}\\[-0.1in]
&    \UpType{1}{2} & (\irrep{10})\irrepsub{495}{1}.(1)\irrepbarsub{66}{H}.\massivefermionpair{(\irrepbar{10})\irrepbar{66}}{(\irrep{10})\irrep{66}}.(\irrep{5})\irrepbarsub{495}{H}.\massivefermionpair{(\irrep{10})\irrep{66}}{(\irrepbar{10})\irrepbar{66}}.(1)\irrepsub{792}{H}.(\irrep{10})\irrepbarsub{220}{2}\\[-0.1in]
&    \UpType{1}{2} & (\irrep{10})\irrepsub{495}{1}.(1)\irrepbarsub{66}{H}.\massivefermionpair{(\irrepbar{10})\irrepbar{66}}{(\irrep{10})\irrep{66}}.(\irrep{5})\irrepbarsub{495}{H}.(\massivefermionpair{\irrep{10})\irrep{66}}{(\irrepbar{10})\irrepbar{66}}.(24)\irrepsub{5148}{H}.(\irrep{10})\irrepbarsub{220}{2}\\[-0.1in]
&    \UpType{2}{1} & (\irrep{10})\irrepbarsub{220}{2}.(1)\irrepsub{792}{H}.\massivefermionpair{(\irrepbar{10})\irrepbar{66}}{(\irrep{10})\irrep{66}}.(\irrep{5})\irrepbarsub{495}{H}.\massivefermionpair{(\irrep{10})\irrep{66}}{(\irrepbar{10})\irrepbar{66}}.(1)\irrepbarsub{66}{H}.(\irrep{10})\irrepsub{495}{1}\\[-0.1in]
&    \UpType{2}{1} & (\irrep{10})\irrepbarsub{220}{2}.(24)\irrepsub{5148}{H}.\massivefermionpair{(\irrepbar{10})\irrepbar{66}}{(\irrep{10})\irrep{66}}.(\irrep{5})\irrepbarsub{495}{H}.\massivefermionpair{(\irrep{10})\irrep{66}}{(\irrepbar{10})\irrepbar{66}}.(1)\irrepbarsub{66}{H}.(\irrep{10})\irrepsub{495}{1}\\[-0.1in]
&    \UpType{2}{2} & (\irrep{10})\irrepbarsub{220}{2}.(1)\irrepsub{792}{H}.\massivefermionpair{(\irrepbar{10})\irrepbar{66}}{(\irrep{10})\irrep{66}}.(\irrep{5})\irrepbarsub{495}{H}.\massivefermionpair{(\irrep{10})\irrep{66}}{(\irrepbar{10})\irrepbar{66}}.(1)\irrepsub{792}{H}.(\irrep{10})\irrepbarsub{220}{2}\\[-0.1in]
&    \UpType{2}{2} & (\irrep{10})\irrepbarsub{220}{2}.(1)\irrepsub{792}{H}.\massivefermionpair{(\irrepbar{10})\irrepbar{66}}{(\irrep{10})\irrep{66}}.(\irrep{5})\irrepbarsub{495}{H}.\massivefermionpair{(\irrep{10})\irrep{66}}{(\irrepbar{10})\irrepbar{66}}.(24)\irrepsub{5148}{H}.(\irrep{10})\irrepbarsub{220}{2}\\[-0.1in]
&    \UpType{2}{2} & (\irrep{10})\irrepbarsub{220}{2}.(24)\irrepsub{5148}{H}.\massivefermionpair{(\irrepbar{10})\irrepbar{66}}{(\irrep{10})\irrep{66}}.(\irrep{5})\irrepbarsub{495}{H}.\massivefermionpair{(\irrep{10})\irrep{66}}{(\irrepbar{10})\irrepbar{66}}.(1)\irrepsub{792}{H}.(\irrep{10})\irrepbarsub{220}{2}\\[-0.1in]
&    \UpType{2}{2} & (\irrep{10})\irrepbarsub{220}{2}.(24)\irrepsub{5148}{H}.\massivefermionpair{(\irrepbar{10})\irrepbar{66}}{(\irrep{10})\irrep{66}}.(\irrep{5})\irrepbarsub{495}{H}.\massivefermionpair{(\irrep{10})\irrep{66}}{(\irrepbar{10})\irrepbar{66}}.(24)\irrepsub{5148}{H}.(\irrep{10})\irrepbarsub{220}{2}\\
\hline
\multicolumn{3}{l}{\bf Leading Down-Type Diagrams:}\\[-0.1in]
\textbf{Dim 4:} \\[-0.1in]
&    \DownType{3}{2} & (\irrep{10})\irrep{66_3}.(\irrepbar{5})\irrepbarsub{12}{H}.(\irrepbar{5})\irrepbar{12_2}\\[-0.1in]
&    \DownType{3}{3} & (\irrep{10})\irrep{66_3}.(\irrepbar{5})\irrepbarsub{12}{H}.(\irrepbar{5})\irrepbar{12_3}\\[-0.1in]
\textbf{Dim 5:} \\[-0.1in]
&    \DownType{1}{2} & (\irrep{10})\irrepsub{495}{1}.(1)\irrepbarsub{66}{H}.\massivefermionpair{(\irrepbar{10})\irrepbar{66}}{(\irrep{10})\irrep{66}}.(\irrepbar{5})\irrepbarsub{12}{H}.(\irrepbar{5})\irrepbarsub{12}{2}\\[-0.1in]
&    \DownType{2}{1} & (\irrep{10})\irrepbarsub{220}{2}.(\irrepbar{5})\irrepbarsub{12}{H}.\massivefermionpair{(\irrepbar{5})\irrep{495}}{(\irrep{5})\irrepbar{495}}.(\irrep{1})\irrepbarsub{792}{H}.(\irrepbar{5})\irrepbarsub{220}{1}\\[-0.1in]
&    \DownType{1}{3} & (\irrep{10})\irrepsub{495}{1}.(\irrep{1})\irrepbarsub{66}{H}.\massivefermionpair{(\irrepbar{10})\irrepbar{66}}{(\irrep{10})\irrep{66}}.(\irrepbar{5})\irrepbarsub{12}{H}.(\irrepbar{5})\irrepbarsub{12}{3}\\[-0.1in]
&    \DownType{3}{1} & (\irrep{10})\irrepsub{66}{3}.(\irrepbar{5})\irrepbarsub{12}{H}.\massivefermionpair{(\irrepbar{5})\irrepbar{12}}{(\irrep{5})\irrep{12}}.(\irrep{1})\irrepsub{66}{H}.(\irrepbar{5})\irrepbarsub{220}{1}\\[-0.1in]
&    \DownType{2}{2} & (\irrep{10})\irrepbarsub{220}{2}.(\irrepbar{5})\irrepbarsub{12}{H}.\massivefermionpair{(\irrepbar{5})\irrep{495}}{(\irrep{5})\irrepbar{495}}.(\irrep{1})\irrepsub{792}{H}.(\irrepbar{5})\irrepbarsub{12}{2}\\[-0.1in]
&    \DownType{2}{2} & (\irrep{10})\irrepbarsub{220}{2}.(\irrep{1})\irrepsub{792}{H}.\massivefermionpair{(\irrepbar{10})\irrepbar{66}}{(\irrep{10})\irrep{66}}.(\irrepbar{5})\irrepbarsub{12}{H}.(\irrepbar{5})\irrepbarsub{12}{2}\\[-0.1in]
&    \DownType{2}{2} & (\irrep{10})\irrepbarsub{220}{2}.(\irrepbar{5})\irrepbarsub{12}{H}.\massivefermionpair{(\irrepbar{5})\irrep{495}}{(\irrep{5})\irrepbar{495}}.(\irrep{24})\irrepsub{5148}{H}.(\irrepbar{5})\irrepbarsub{12}{2}\\[-0.1in]
&    \DownType{2}{2} & (\irrep{10})\irrepbarsub{220}{2}.(\irrep{24})\irrepsub{5148}{H}.\massivefermionpair{(\irrepbar{10})\irrepbar{66}}{(\irrep{10})\irrep{66}}.(\irrepbar{5})\irrepbarsub{12}{H}.(\irrepbar{5})\irrepbarsub{12}{2}\\[-0.1in]
&    \DownType{2}{3} & (\irrep{10})\irrepbarsub{220}{2}.(\irrepbar{5})\irrepbarsub{12}{H}.\massivefermionpair{(\irrepbar{5})\irrep{495}}{(\irrep{5})\irrepbar{495}}.(\irrep{1})\irrepsub{792}{H}.(\irrepbar{5})\irrepbarsub{12}{3}\\[-0.1in]
&    \DownType{2}{3} & (\irrep{10})\irrepbarsub{220}{2}.(\irrep{1})\irrepsub{792}{H}.\massivefermionpair{(\irrepbar{10})\irrepbar{66}}{(\irrep{10})\irrep{66}}.(\irrepbar{5})\irrepbarsub{12}{H}.(\irrepbar{5})\irrepbarsub{12}{3}\\[-0.1in]
&    \DownType{2}{3} & (\irrep{10})\irrepbarsub{220}{2}.(\irrepbar{5})\irrepbarsub{12}{H}.\massivefermionpair{(\irrepbar{5})\irrep{495}}{(\irrep{5})\irrepbar{495}}.(\irrep{24})\irrepsub{5148}{H}.(\irrepbar{5})\irrepbarsub{12}{3}\\[-0.1in]
&    \DownType{2}{3} & (\irrep{10})\irrepbarsub{220}{2}.(\irrep{24})\irrepsub{5148}{H}.\massivefermionpair{(\irrepbar{10})\irrepbar{66}}{(\irrep{10})\irrep{66}}.(\irrepbar{5})\irrepbarsub{12}{H}.(\irrepbar{5})\irrepbarsub{12}{3}\\[-0.1in]
\textbf{Dim 6:} \\[-0.1in]
&    \DownType{1}{1} & (\irrep{10})\irrepsub{495}{1}.(\irrep{1})\irrepbarsub{66}{H}.\massivefermionpair{(\irrepbar{10})\irrepbar{66}}{(\irrep{10})\irrep{66}}.(\irrepbar{5})\irrepbarsub{12}{H}.\massivefermionpair{(\irrepbar{5})\irrepbar{12}}{(\irrep{5})\irrep{12}}.(\irrep{1})\irrepsub{66}{H}.(\irrepbar{5})\irrepbarsub{220}{1}\\[-0.1in]
&    \DownType{1}{1} & (\irrep{10})\irrepsub{495}.(\irrep{1})\irrepsub{792}{H}.\massivefermionpair{(\irrepbar{10})\irrep{220}}{(\irrep{10})\irrepbar{220}}.(\irrepbar{5})\irrepbarsub{12}{H}.\massivefermionpair{(\irrepbar{5})\irrep{495}}{(\irrep{5})\irrepbar{495}}.(\irrep{1})\irrepbarsub{792}{H}.(\irrepbar{5})\irrepbarsub{220}{1}\\
\hline
\end{tabular}
\end{center}     
\end{table}

%% file: matrixelementtablecont10C.tex
\begin{table}
\begin{flushleft}
\renewcommand\boldirrep\relax
\setlength{\arraycolsep}{1pt}
\begin{tabular*}{6.5in}{lll}\\[-14pt]
\hline
\multicolumn{3}{l}{\bf Leading Dirac Neutrino Diagrams:}\\[-0.1in]
\textbf{Dim 5:} \\[-0.1in]
&    \Dirac{1}{1} & (\irrepbar{5})\irrepbarsub{220}{1}.(5)\irrepbarsub{495}{H}.\massivefermionpair{(\irrep{1})\irrepbar{792}}{(\irrep{1})\irrep{792}}.(\irrep{1})\irrepbarsub{792}{H}.(\irrep{1})\irrepsub{1}{1} \hfill\\[-0.1in]
&    \Dirac{1}{1} & (\irrepbar{5})\irrepbarsub{220}{1}.(1)\irrepbarsub{792}{H}.\massivefermionpair{(\irrep{5})\irrepbar{495}}{(\irrepbar{5})\irrep{495}}.(\irrep{5})\irrepbarsub{495}{H}.(1)\irrepsub{1}{1}\\[-0.1in]
&    \Dirac{1}{2} & (\irrepbar{5})\irrepbarsub{220}{1}.(\irrep{5})\irrepbarsub{495}{H}.\massivefermionpair{(\irrep{1})\irrepbar{792}}{(\irrep{1})\irrep{792}}.(\irrep{1})\irrepbarsub{792}{H}.(\irrep{1})\irrepsub{1}{2}\\[-0.1in]
&    \Dirac{1}{2} & (\irrepbar{5})\irrepbarsub{220}{1}.(\irrep{1})\irrepbarsub{792}{H}.\massivefermionpair{(\irrep{5})\irrepbar{495}}{(\irrepbar{5})\irrep{495}}.(\irrep{5})\irrepbarsub{495}{H}.(\irrep{1})\irrepsub{1}{2}\\[-0.1in]
&    \Dirac{2}{1} & (\irrepbar{5})\irrepbarsub{12}{2}.(\irrep{5})\irrepbarsub{495}{H}.\massivefermionpair{(\irrep{1})\irrep{792}}{(\irrep{1})\irrepbar{792}}.(\irrep{1})\irrepsub{792}{H}.(1)\irrepsub{1}{1}\\[-0.1in]
&    \Dirac{2}{1} & (\irrepbar{5})\irrepbarsub{12}{2}.(\irrep{1})\irrepsub{792}{H}.\massivefermionpair{(\irrep{5})\irrepbar{495}}{(\irrepbar{5})\irrep{495}}.(\irrep{5})\irrepbarsub{495}{H}.(\irrep{1})\irrepsub{1}{1}\\[-0.1in]
&    \Dirac{2}{1} & (\irrepbar{5})\irrepbarsub{12}{2}.(\irrep{24})\irrepsub{5148}{H}.\massivefermionpair{(\irrep{5})\irrepbar{495}}{(\irrepbar{5})\irrep{495}}.(\irrep{5})\irrepbarsub{495}{H}.(\irrep{1})\irrepsub{1}{1}\\[-0.1in]
&    \Dirac{1}{3} & (\irrepbar{5})\irrepbarsub{220}{1}.(\irrep{5})\irrepbarsub{495}{H}.\massivefermionpair{(\irrep{1})\irrepbar{792}}{(\irrep{1})\irrep{792}}.(\irrep{1})\irrepbarsub{792}{H}.(\irrep{1})\irrepsub{1}{3}\\[-0.1in]
&    \Dirac{1}{3} & (\irrepbar{5})\irrepbarsub{220}{1}.(\irrep{1})\irrepbarsub{792}{H}.\massivefermionpair{(\irrep{5})\irrepbar{495}}{(\irrepbar{5})\irrep{495}}.(\irrep{5})\irrepbarsub{495}{H}.(1)\irrepsub{1}{3}\\[-0.1in]
&    \Dirac{3}{1} & (\irrepbar{5})\irrepbarsub{12}{3}.(\irrep{5})\irrepbarsub{495}{H}.\massivefermionpair{(\irrep{1})\irrep{792}}{(\irrep{1})\irrepbar{792}}.(\irrep{1})\irrepsub{792}{H}.(1)\irrepsub{1}{1}\\[-0.1in]
&    \Dirac{3}{1} & (\irrepbar{5})\irrepbarsub{12}{3}.(1)\irrepsub{792}{H}.\massivefermionpair{(\irrep{5})\irrepbar{495}}{(\irrepbar{5})\irrep{495}}.(\irrep{5})\irrepbarsub{495}{H}.(1)\irrepsub{1}{1}\\[-0.1in]
&    \Dirac{3}{1} & (\irrepbar{5})\irrepbarsub{12}{3}.(\irrep{24})\irrepsub{5148}{H}.\massivefermionpair{(\irrep{5})\irrepbar{495}}{(\irrepbar{5})\irrep{495}}.(\irrep{5})\irrepbarsub{495}{H}.(1)\irrepsub{1}{1}\\[-0.1in]
&    \Dirac{2}{2} & (\irrepbar{5})\irrepbarsub{12}{2}.(\irrep{5})\irrepbarsub{495}{H}.\massivefermionpair{(\irrep{1})\irrep{792}}{(\irrep{1})\irrepbar{792}}.(\irrep{1})\irrepsub{792}{H}.(1)\irrepsub{1}{2}\\[-0.1in]
&    \Dirac{2}{2} & (\irrepbar{5})\irrepbarsub{12}{2}.(\irrep{1})\irrepsub{792}{H}.\massivefermionpair{(\irrep{5})\irrepbar{495}}{(\irrepbar{5})\irrep{495}}.(\irrep{5})\irrepbarsub{495}{H}.(1)\irrepsub{1}{2}\\[-0.1in]
&    \Dirac{2}{2} & (\irrepbar{5})\irrepbarsub{12}{2}.(24)\irrepsub{5148}{H}.\massivefermionpair{(\irrep{5})\irrepbar{495}}{(\irrepbar{5})\irrep{495}}.(\irrep{5})\irrepbarsub{495}{H}.(1)\irrepsub{1}{2}\\[-0.1in]
&    \Dirac{2}{3} & (\irrepbar{5})\irrepbarsub{12}{2}.(\irrep{5})\irrepbarsub{495}{H}.\massivefermionpair{(\irrep{1})\irrep{792}}{(\irrep{1})\irrepbar{792}}.(\irrep{1})\irrepsub{792}{H}.(1)\irrepsub{1}{3}\\[-0.1in]
&    \Dirac{2}{3} & (\irrepbar{5})\irrepbarsub{12}{2}.(1)\irrepsub{792}{H}.\massivefermionpair{(\irrep{5})\irrepbar{495}}{(\irrepbar{5})\irrep{495}}.(\irrep{5})\irrepbarsub{495}{H}.(1)\irrepsub{1}{3}\\[-0.1in]
&    \Dirac{2}{3} & (\irrepbar{5})\irrepbarsub{12}{2}.(24)\irrepsub{5148}{H}.\massivefermionpair{(\irrep{5})\irrepbar{495}}{(\irrepbar{5})\irrep{495}}.(\irrep{5})\irrepbarsub{495}{H}.(1)\irrepsub{1}{3}\\[-0.1in]
&    \Dirac{3}{2} & (\irrepbar{5})\irrepbarsub{12}{3}.(\irrep{5})\irrepbarsub{495}{H}.\massivefermionpair{(\irrep{1})\irrep{792}}{(\irrep{1})\irrepbar{792}}.(\irrep{1})\irrepsub{792}{H}.(1)\irrepsub{1}{2}\\[-0.1in]
&    \Dirac{3}{2} & (\irrepbar{5})\irrepbarsub{12}{3}.(\irrep{1})\irrepsub{792}{H}.\massivefermionpair{(\irrep{5})\irrepbar{495}}{(\irrepbar{5})\irrep{495}}.(\irrep{5})\irrepbarsub{495}{H}.(1)\irrepsub{1}{2}\\[-0.1in]
&    \Dirac{3}{2} & (\irrepbar{5})\irrepbarsub{12}{3}.(\irrep{24})\irrepsub{5148}{H}.\massivefermionpair{(\irrep{5})\irrepbar{495}}{(\irrepbar{5})\irrep{495}}.(\irrep{5})\irrepbarsub{495}{H}.(1)\irrepsub{1}{2}\\[-0.1in]
&    \Dirac{3}{3} & (\irrepbar{5})\irrepbarsub{12}{3}.(5)\irrepbarsub{495}{H}.\massivefermionpair{(\irrep{1})\irrep{792}}{(\irrep{1})\irrepbar{792}}.(\irrep{1})\irrepsub{792}{H}.(1)\irrepsub{1}{3}\\[-0.1in]
&    \Dirac{3}{3} & (\irrepbar{5})\irrepbarsub{12}{3}.(\irrep{1})\irrepsub{792}{H}.\massivefermionpair{(\irrep{5})\irrepbar{495}}{(\irrepbar{5})\irrep{495}}.(\irrep{5})\irrepbarsub{495}{H}.(1)\irrepsub{1}{3}\\[-0.1in]
&    \Dirac{3}{3} & (\irrepbar{5})\irrepbarsub{12}{3}.(\irrep{24})\irrepsub{5148}{H}.\massivefermionpair{(\irrep{5})\irrepbar{495}}{(\irrepbar{5})\irrep{495}}.(\irrep{5})\irrepbarsub{495}{H}.(1)\irrepsub{1}{3}\\
\hline
\multicolumn{3}{l}{\bf Leading Majorana Neutrino Diagrams:}\hfill\\[-0.1in]
\textbf{Dim 4:} \\[-0.1in]
&    \Majorana{1}{1} & (\irrep{1})\irrepsub{1}{1}.(1)\irrepsub{1}{H}.(1)\irrepsub{1}{1}\\[-0.1in]
&    \Majorana{1}{2} & (\irrep{1})\irrepsub{1}{1}.(1)\irrepsub{1}{H}.(1)\irrepsub{1}{2}\\[-0.1in]
&    \Majorana{2}{1} & (\irrep{1})\irrepsub{1}{2}.(1)\irrepsub{1}{H}.(1)\irrepsub{1}{1}\\[-0.1in]
&    \Majorana{2}{2} & (\irrep{1})\irrepsub{1}{2}.(1)\irrepsub{1}{H}.(1)\irrepsub{1}{2}\\[-0.1in]
&    \Majorana{1}{3} & (\irrep{1})\irrepsub{1}{1}.(1)\irrepsub{1}{H}.(1)\irrepsub{1}{3}\\[-0.1in]
&    \Majorana{3}{1} & (\irrep{1})\irrepsub{1}{3}.(1)\irrepsub{1}{H}.(1)\irrepsub{1}{1}\\[-0.1in]
&    \Majorana{2}{3} & (\irrep{1})\irrepsub{1}{2}.(1)\irrepsub{1}{H}.(1)\irrepsub{1}{3}\\[-0.1in]
&    \Majorana{3}{2} & (\irrep{1})\irrepsub{1}{3}.(1)\irrepsub{1}{H}.(1)\irrepsub{1}{2}\\[-0.1in]
&    \Majorana{3}{3} & (\irrep{1})\irrepsub{1}{3}.(1)\irrepsub{1}{H}.(1)\irrepsub{1}{3}\\
\hline\hline
\end{tabular*}
\caption{\label{tab:massmatrixelementsforSpecialCase}Yukawa diagrams for the up
 and down quark and Dirac neutrino matrices, as well as the righthanded 
 Majorana neutrino diagrams, for the special model singled out in Sect. IV C.  
 which belongs to the {\bf IC}6 class of models of Table II.  The diagrams for 
 the charged leptons are the transpose of the down quark diagrams.}
\end{flushleft}
\end{table}